\documentclass[aps,pre,twocolumn,superscriptaddress,babel]{revtex4-1}

\usepackage{mathtools,overpic,float}
\usepackage{amsmath,amssymb,graphicx,color}
\usepackage{bm}
\usepackage{bbm}

\def\be{\begin{equation}}
\def\ee#1{\label{#1}\end{equation}}
\def\a{\alpha}
\def\b{\beta}
\def\g{\gamma}
\def\G{\Gamma}
\def\d{\delta}

\def\e{\epsilon}
\def\z{\zeta}
\def\h{\eta}
\def\th{\theta}

\def\k{\kappa}
\def\l{\lambda}
\def\La{\Lambda}

\def\n{\nu}

\def\X{\Xi}
\def\p{\pi}
\def\r{\rho}
\def\s{\sigma}
\def\t{\tau}

\def\F{\Phi}
\def\c{\chi}
\def\ps{\psi}
\def\o{\omega}

\def\i{\int}

\def\bM{{\mathbf M}}

\def\bfz{\mbox{\boldmath{$\z$}}}

\def\Tr{\mbox{Tr}}

\begin{document}
\title{Measurement driven single temperature engine}
\author{ Xuehao Ding}
\affiliation{School of Physics, Peking University, Beijing, P.R. China, 100871}
\author{Juyeon Yi}
\affiliation{Department of Physics, Pusan National University, Busan 46241, Republic of Korea}
\author{Yong Woon Kim}
\affiliation{Graduate School of Nanoscience and Technology, Korea Advanced Institute of Science and Technolgy, Deajeon 34141, Republic of Korea}
\author{Peter Talkner}
\affiliation{Institut f\"{u}r Physik, Universit\"{a}t Augsburg, Universit\"{a}tsstra{\ss}e 1, D-86135 Augsburg, Germany}
\date{\today}
\begin{abstract}
A four stroke quantum engine which alternately interacts with a measurement apparatus and a single heat bath is discussed in detail with respect to the average work and heat as well as to the fluctuations of work and heat. The efficiency and the reliability of such an engine with a harmonic oscillator as working substance are analyzed under different conditions such as different speeds of the work strokes, different temperatures of the heat bath and various strengths of the energy supplying measurement. For imperfect thermalization strokes of finite duration also the power of the engine is analyzed. A comparison with a two-temperature Otto engine is provided in the particular case of adiabatic work  and ideal thermalization strokes.     
\end{abstract}
\maketitle
\section{Introduction}
The practical understanding of how to convert heat into useful work, beginning in the 18th century, did not only trigger the industrial revolution but also spurred the development of thermodynamics \cite{Planck}. Motivated by novel experimental techniques operating on molecular or even atomic scales the basic ideas underlying conventional heat engines have been applied to  understand biological processes on cellular and molecular scales in thermodynamical terms  and also been used to design  artificial machines on  microscopic scales \cite{JAP,R,HM,BCSW}.  
The interpretation of a maser in terms of a Carnot cycle was proposed in Refs. \cite{SdB,Ko,BBM}. Microscopic engines performing other thermodynamic cycles such as Otto motors \cite{FK,KR} or Szilard engines \cite{S,QWLSN} were suggested and discussed both for classical and quantum mechanical models.
Szilard-type engines differ from standard heat engines operating between two heat baths at different temperatures in that the energy is extracted from a single heat bath by means of a feedback mechanism agitated by a so-called Maxwell's demon \cite{LR}. Of course, the idea of feedback is much older and has been used to control engines from the early times of James Watt on.
The use of squeezed reservoirs instead of thermal heat baths was first suggested as a genuinely quantum mechanical boosting mechanism in \cite{Sc}  and later experimentally demonstrated \cite{RASSL}.  

A quantum engine in which, as the energy providing stroke, the contact with a hot heat bath is replaced by the measurement of a properly chosen observable was recently proposed \cite{YTK}. 
In this setting the result of the measurement is not further used to control the engine. Rather the back-action of this measurement provides the ``fuel'' of such a device  under suitable conditions.  
In Ref. \cite{EHHA,EJ} selective measurements in combination with feedback control are suggested as the energy source of a Szilard-type quantum engine. A comparison of the non-selective measurement scenario with a selective measurement whose result is used for controlling  the considered engine is analyzed in \cite{YK}.

Even though an engine of molecular size acting according to the laws of quantum mechanics can be expected to display large random deviations from its average behavior, with a few exceptions \cite{RHM,CPF,LKAS},   most of the existing investigations only deal with the behavior of average work and heat and quantities derived therefrom such as efficiency and power. The averages of work and heat are most often determined in terms of respective powers that are obtained by splitting the time rate of change of the average energy of the working substance into parts that are interpreted as power and heat rate \cite{A}. 
However, it is not possible to translate the calculational prescription for the average of work based on power into an operational definition of a fluctuating work \cite{VWT}. 
Therefore we will use the two energy measurement approach as an operational definition of work \cite{K,Tas,TLH,TH} and heat \cite{TCH,CHT,GPM}. This requires to allow for a number of diagnostic energy measurements within each cycle of the considered engine \cite{ZHP}. As a result we not only get access to the average behavior of the engine but also to its random properties described in terms of work and heat. In the case of a truly quantum engine, the obtained results actually will depend on the chosen diagnostic tools as was recently demonstrated for an Otto engine interacting with an auxiliary work deposit  \cite{WVT}.   

The paper is organized as follows. In Section~\ref{fourstrokes} we recapitulate the  idea of the measurement driven engine proposed in \cite{YTK} and specify the four subsequent strokes in combination with the diagnostic energy measurements in Section~\ref{energy}. Each stroke can be characterized by a transition probability between energy eigenstates of the instantaneous Hamiltonians immediately before and after a stroke. 
These transition probabilities are specified for a harmonic oscillator with externally controllable frequency  in Section~\ref{harmosc}  
and utilized in Section~\ref{perf} to discuss  the performance of a measurement engine both with ideal and imperfect thermalization. In the latter case, a weak contact with the heat-bath during a finite amount of time is modeled by a Markovian master equation. A comparison between the measurement engine and a two temperature heat engine is provided in Section~\ref{Otto}. 
A summary and discussions of the main results are presented in Section~\ref{conclusion} followed by Appendices with some technical details.

\section{Four stroke measurement engine} \label{fourstrokes}
We consider an engine consisting of three basic elements: (i) a working substance governed by a Hamiltonian $H(\l)$, and controlled by the change of a time-dependent parameter $\l$, (ii) an  energy source, and (iii) a thermal reservoir. For example, taking as element (i) for the working substance a gas  whose volume is externally controlled by the position of a piston,  as element (ii) a hot heat-bath and as element (iii) the thermal reservoir at a lower temperature one has assembled  the elements of a traditional heat engine such as an Otto engine. 
Here we instead study a quantum system as the working substance, which, for the sake of definiteness, later on will be chosen as a harmonic oscillator whose frequency can externally be controlled. The energy input is provided by the measurement of an observable of the working substance. In order that energy is put into the system it is necessary that this observable does not commute with the Hamiltonian describing the working substance at the instant of the measurement. The outcome of the measurement is ignored; rather the genuinely quantum mechanical back-action of the measurement on the state of the working substance results in a transfer of energy. The thermal reservoir can be at any temperature.
All parameter changes of the working substance are performed while the system is decoupled from the thermal reservoir. The measurement is also taken in a phase during which the system is thermally isolated and the parameter $\l$ is kept constant. To be precise, a complete cycle of the engine consists of four strokes which proceed as follows: The engine starts in a state ${\bf 0}$ of thermal equilibrium at the inverse temperature $\b$; subsequently the working substance is compressed by a change of the parameter from $\l_i$ to $\l_f$ according to a prescribed protocol, leading to an increase of the energy-level distances in the state ${\bf 1}$. After completion of this first work stroke, the working substance is kept at the reached parameter value $\l_f$ and a measurement of the oscillator position is performed transferring the working substance into the state ${\bf 2}$. Out of state ${\bf 2}$ the working substance is expanded back to the initial value of the parameter ending up with $H(\l_i)$ in the state ${\bf 3}$.  In order to fully recover the initial state ${\bf 0}$ the working substance is brought into weak contact with the thermal reservoir acting as a heat-bath at the initial inverse temperature $\b$. A schematic view of a cycle is depicted in Eq. (\ref{cycle}).
\be
{\bf 0} \xRightarrow{\text{WS I\; }} {\bf 1} \xRightarrow{\text{QM}} {\bf 2} 
\xRightarrow{\text{WS II}} {\bf 3}
\xRightarrow{\text{T\;}}{\bf 0'}\:.
\ee{cycle}

As a diagnostic tool projective measurements of the energy are performed at the beginning in the state ${\bf 0}$ and after the completion of each stroke, hence yielding a sequence of four energies $E_0$, $E_1$, $E_2$, $E_3$ constituting the energy profile of a complete engine cycle.  
These energies hence coincide with one of the eigenvalues of either Hamiltonian $H(\l_i) = \sum_k \e_k(\l_i) P_k(\l_i)$ or $H(\l_f)=\sum_k \e_k(\l_f) P_k(\l_f)$ governing the working substance in the states $\mathbf{0}$, $\mathbf{3}$,  and $\mathbf{1}$, $\mathbf{2}$, respectively. Here, the operators $P_k(\l_i)$ and $P_k(\l_f)$ project onto the eigenspaces corresponding to the respective eigenvalues $\e_k(\l_i)$ and $\e_k(\l_f)$. Therefore, one obtains $E_0,E_3, \in \{\e_k(\l_i),k=1,2,\ldots \}$ and $E_1,E_2 \in \{\e_k(\l_f),k=1,2,\ldots \}$. 

Knowing the measured energies one can determine the amounts of work, $W_I$ and $W_{II}$ performed on the working substance during the two work strokes WS I and WS II, respectively as
\be
W_I= E_1-E_0\:,
\ee{WI}
and
\be
W_{II} = E_3 - E_2\:.
\ee{WII}
Accordingly the energy change $E_{\mathcal{M}}$ caused by the measurement is given by
\be
E_{\mathcal{M}} =E_2 -E_1\:,
\ee{EM}
and, finally, the heat that is exchanged with the thermal reservoir, follows as
\be
Q=E_{0'} -E_3\:.
\ee{Q}
where $0'$ is the initial state of the next cycle.
Here we adopted the sign  convention that a positive change of energy, be it work, measurement energy or heat, corresponds to an increase of the energy of the system.

General expressions for the averages of the total work $\langle W\rangle \equiv \langle W_I \rangle + \langle W_{II} \rangle$ and the supplied energy $\langle E_{\mathcal{M}} \rangle$ are given in \cite{YTK} for adiabatic work strokes and the large class of minimally disturbing generalized measurements, which also includes projective measurements. In Ref.~\cite{YTK} the two adiabatic work strokes form a pair of mutually time-reversed processes that proceed infinitely slowly such that transitions between different instantaneous energy levels do not occur.
Knowing the total work and the supplied energy, one can determine the efficiency $\h$ as
\be
\h = -\frac{\langle W \rangle}{\langle E_{\mathcal{M}}\rangle}\:.
\ee{eta}  
For the special situation of uniform adiabatic compression and expansion of the working substance where all energy differences between corresponding pairs of energy eigenvalues  expand and shrink at a constant compression rate $\g$, i.e.  $\e_k(\l_f) - \e_l(\l_f) =\g(\e_k(\l_i) - \e_l(\l_i))$, \cite{note}  the efficiency takes the universal form
\be
\h = 1 - \g^{-1}\:.
\ee{hg}  

Before we 
consider the averages and fluctuations of the total work as well as of the supplied energy for the particular example of a harmonic oscillator with time-dependent frequency as working substance, we  give a general description of the energy profile of a cycle in terms of the joint distribution of the five energy values introduced above.
\section{Energy profile} \label{energy}
The joint distribution of the energies $E_0,\ldots E_{0'} $ is given by
\be
\begin{split}
p(n',l,k,m,n) &=     T_\b(n',l) T_{\text{WS\:II}}(l,k) T_M(k,m)\\ 
&\quad \times T_{\text{WS\:I}}(m,n) p_\b(n)\:,
\end{split}
\ee{p5}
where $n,m,k,l,n'$ label the energy eigenvalues in the states ${\bf 0}, {\bf 1}, {\bf 2}, {\bf 3}, {\bf 0'}$. Further,
\be
p_\b(n) = Z^{-1} e^{-\b e_n(\l_i)}\:, \quad Z = \sum_n e^{-\b e_n(\l_i)}  
\ee{p0}  
denotes the thermal distribution in the initial state ${\bf 0}$ and  $T_{\text{WS\:I}}(m,n)$, $T_M(m,k)$, $T_{\text{WS\:II}}(l,k)$, $T_\b(n',l)$ the transition probabilities between the according energy levels induced by the compression (WS\:I), measurement (QM), expansion (WS\:II) and thermalization (T) strokes, respectively. As transition probabilities they satisfy $0\leq T_X(m,n) \leq 1$ and $\sum_m T_X(m,n) =1$ for $X=TW\:I, M, TW\:II, \b$ as can be seen from the explicit expressions given below.  

\subsection{Work strokes} \label{ws}
For the sake of simplicity we assume that the two work-strokes are mutually time-reversed so that 
\be
T_{\text{WS\:II}}(l,k) = T_{\text{WS\:I}}(k,l) \equiv T(k,l)\:,
\ee{TexpTcomp}
yielding for a compression protocol $\La =\{\l(t)|t_i<t<t_f \}$ with $\l(t_i) =\l_i$ and $\l(t_f) = \l_f$ the transition probability
\be
T(m,n) =\left |\langle m;\l_f| \mathcal{T} \exp \left \{ -\frac{i}{\hbar} \i_{t_i}^{t_f} ds H(\l(s)) \right \}| n; \l_i  \rangle \right |^2\:,
\ee{T}     
where $\mathcal{T}$ denotes chronological time ordering; the states $|n;\l_i \rangle$
and $|m;\l_f \rangle$   are eigenvectors of the Hamiltonians $H(\l_i)$ and $H(\l_f)$, respectively. Further, we assume that there are no degeneracies of the energy eigenvalues at any time of the protocol, and, hence, no crossings of energy levels occur; therefore the previously introduced projection operators become $P_n(\l) =|n;\l\rangle \langle n;\l|$ with $\l=\l_i,\l_f$. 

In the particular case of an adiabatic, meaning an infinitely slow, process there are no transitions between the different energy eigenstates and hence
\be
T^{\text{ad}}(k,l) =\d_{k,l}\:,
\ee{Tad}
where $\d_{k,l}$ denotes the Kronecker delta symbol.

 \subsection{Measurement stroke} \label{M}
We consider here a non-selective minimally disturbing measurement \cite{WM} of an observable that does not commute with the Hamiltonian $H(\l_f)$ governing the working substance at the instant of the measurement. The action of a non-selective minimally disturbing measurement is characterized by an operation $\F_{\mathcal{M}}$, i.e. a linear, completely positive map that  transforms the density matrix $\r$ prior to the measurement into the post-measurement density matrix  $\r' = \F_{\mathcal{M}}(\r)$ \cite{WM}. Because $\F_{\mathcal{M}}$ describes a non-selective measurement, it conserves the trace, $\Tr \F_{\mathcal{M}}(\r) = \Tr \r$ for all trace-class operators $\r$, and, as a minimally disturbing measurement, it is unital, $\F^*_{\mathcal{M}} = \F_{\mathcal{M}}$, with the dual operation $\F^*_{\mathcal{M}}$ defined by $\Tr u \F_{\mathcal{M}}(\r) = \Tr \F^*_{\mathcal{M}}(u) \r$ for all bounded operators $u$ and all trace class operators $\r$.   The transition probability between states labeled by the indices $m$ and $k$ resulting from the measurement operation $\F_{\mathcal{M}}$ can be expressed as
\be
\begin{split}
T_{\mathcal{M}}(k,m) &= \Tr P_k(\l_f) \F_{\mathcal{M}}(P_m(\l_f))\\
& = \langle k;\l_f|\F_{\mathcal{M}}(P_m(\l_f))|k;\l_f \rangle \:,
\end{split}
\ee{TM}
where the second equality is obtained for non-degenerate energy eigenvalues.
As a consequence of $\F_{\mathcal{M}}$ being unital it follows that $T_{\mathcal{M}}(k,m)= T_{\mathcal{M}}(m,k)$ is symmetric in its indices. It further implies that the average energy change $\langle E_{\mathcal{M}} \rangle =\sum_{n,m} (\e_n(\l_f) - \e_m(\l_f)) T_{\mathcal{M}}(n,m) p(m)$ induced by such a measurement is positive provided that the energy distribution $p(n) = \langle n; \l_f| \r |n; \l_f \rangle$ decreases with increasing energy $\e_n(\l_f)$ \cite{YTK}. This follows from rewriting the expression for the average with the help of the symmetry of the transition probability as $\langle E_{\mathcal{M}} \rangle = (1/2) \sum_{m,n} (\e_n(\l_f) - \e_m(\l_f)) (p(m) -p(n)) T_{\mathcal{M}}(n,m) \geq 0$. Hence, on average, a non-selective, minimally disturbing measurement cannot extract energy on average out of a system whose  population decreases with increasing energy.       
\subsection{Thermalization} \label{ts}
To achieve thermalization in the last stroke, the working substance has to be brought into week contact with a heat-bath at the required inverse temperature $\b$ for a sufficiently long time. Formally, the result of this process can be described by an operation $\F_\b$ that maps every normalized density matrix $\r$ on the Gibbs state $\r_\b = Z^{-1} e^{- \b H(\l_i)}$, $\F_\b(\r) = \r_\b  \Tr \r$. Hence one obtains for the transition probability 
\be
\begin{split}
T_\b(n',l) &= \Tr P_{n'} \F_\b(P_l(\l_i))\\ 
&= p_\b(n')\:,
\end{split}
\ee{Tb}
where $p_\b(n')$ is the energy distribution of the initial  equilibrium state. When the contact with the heat-bath is maintained only over a finite time, the thermalization will no longer be perfect and the transition probability will also depend on the  state $l$. 
We assume that the influence of the heat-bath can be described by a master equation \cite{BP} of the form $\dot{\r} = \La \r$ with a dissipative Liouvillian $\La$, which has $\r_\b$ as the unique eigenstate belonging to the eigenvalue $0$, hence satisfying $\La \r_\b =0$. The operation  generated by a contact of duration $\th$ between the system and the reservoir then becomes $\F_{\b,\th}(\r) = e^{\La \th} \r$.
In the limit $\th \to \infty$ one recovers the ideal thermalization operation $\F_\b$. 

For both finite-time approximate and infinite-time exact thermalization the interaction between the working substance and the thermal reservoir has to be small because the validity of  Markovian master equations is restricted to weak coupling \cite{T}; at finite coupling the asymptotically reached density matrix is known to deviate from the Gibbs state $\r_\b$ \cite{GWT}.          
\section{Harmonic oscillator as working substance}\label{harmosc}
\begin{figure}
\includegraphics[width=0.45\textwidth]{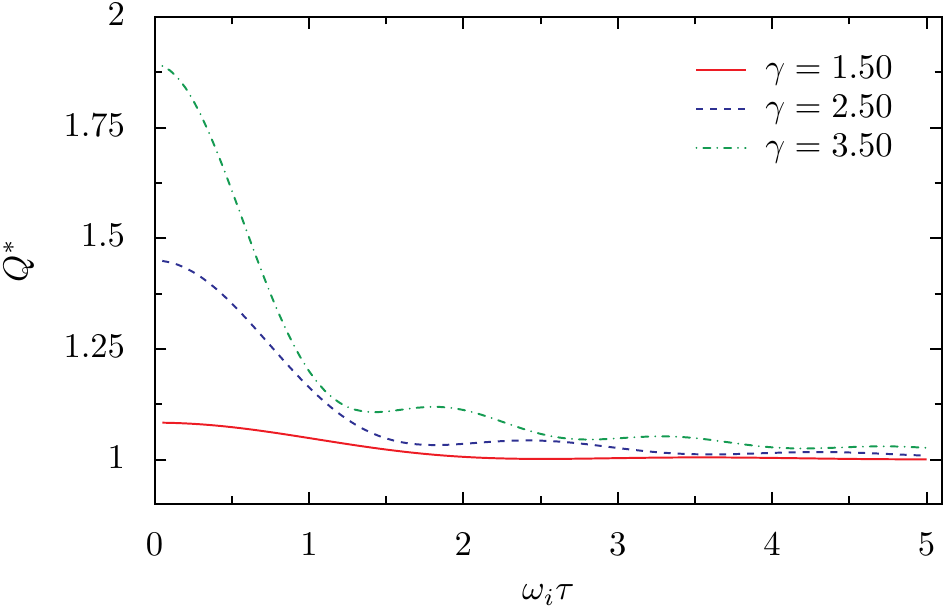}
\caption{
The Husimi parameter $Q^*$ is displayed as a function of the protocol duration $\t$ in units of the initial inverse frequency $\o_i$ for a linear variation of $\o^2(t)$ with different compression ratios $\gamma$. For sudden quenches, i.e. for $\t=0$, the Husimi parameter is largest compared to slower protocols. It approaches the adiabatic value $1$ in an almost monotonic decay which is superimposed by small oscillations.}
\label{hus}
\end{figure}

We consider here a harmonic oscillator of mass $m$ with an externally variable frequency $\o(t)$ as the working substance. Its dynamics is therefore governed by the time-dependent Hamiltonian  $H(\o(t))$ given by
\be
H(\o(t))= \frac{p^2}{2m} + \frac{1}{2} m \o^2(t) q^2\:.
\ee{Ht}
The time-dependent frequency $\o(t)$ plays the role of the work parameter. It changes according to a protocol lasting the time $\t$ from $\o(0) =\o_i$ to $\o(\t) = \o_f$ causing transitions between the eigenstates $|n;\o_i\rangle$ and
 $|m;\o_f\rangle$ of the initial and final Hamiltonians $H(\o_i)$ and $H(\o_f)$, respectively. The transition probability (\ref{T}) can be explicitly expressed as \cite{H,DAL}
\be
T(m,n) = \left \{ \begin{array}{ll}
\frac{2^{1/2}}{(Q^*+1)^{1/2}} \left ( \frac{Q^*-1}{Q^*+1} \right )^{(m+n)/2}&\\
\quad \times \frac{\G((m+1)/2) \G((n+1)/2)}{\p \G(m/2 +1)\G(n/2+1)}&\quad m,n:\;\\
\quad \times {}_2F_1\left (-\frac{m}{2},-\frac{n}{2};\frac{1}{2};\frac{2}{1-Q^*} \right)^2 &\hspace{3mm} \text{even}\\*[3mm]
\frac{2^{7/2}}{(Q^*+1)^{3/2}} \left (\frac{Q^*-1}{Q^*+1} \right )^{(m+n)/2-1}&\\ \quad \times \frac{\G(m/2 +1) \G(n/2+1)}{\p \G((m+1)/2) \G((n+1)/2)}& \quad m,n:\;\\
\quad \times \:{}_2F_1 \left (\frac{1-m}{2},\frac{1-n}{2};\frac{3}{2}; \frac{2}{1-Q^*} \right )^2 &\quad  \text{odd}\\*[3mm]
0  &\quad \text{else}\:,\\
\end{array}
\right .
\ee{Tmn}
where ${}_2F_1(a,b;c;z)$ is a hypergeometric function \cite{AS}. Here the Husimi parameter $Q^*$ contains all information about the frequency protocol.
It is given by~\cite{H}
\be
\begin{split}
Q^*&=\frac{1}{2 \o_i \o_f}\\
&\quad \times \left \{ \o^2_i \left [\o^2_f X^2(\t) + \dot{X}^2(\t) \right ] + \o^2_f Y^2(\t) + \dot{Y}^2(\t) \right \}\:,
\end{split}
\ee{q}
where the functions $X(t)$ and $Y(t)$ are solutions of the equation of motion of a classical harmonic oscillator $\ddot{x}(t) +\o^2(t) x(t) =0$ with the initial conditions at $t_i=0$, $X(0) =0$, $\dot{X}(0) =1$ and $Y(0) =1$, $\dot{Y}(0) =0$. The protocol ends at $t_f =\t$ at the compression ratio $\g = \o_f/\o_i$. 
The Husimi parameter assumes the value 1 for an infinitely slow protocol and reaches the maximal value $Q^*_{\text{quench}} =(\o^2_i+\o^2_f)/(2\o_i \o_f) = (\g + \g^{-1})/2$ for a sudden quench.
Figure~\ref{hus} exemplifies the dependence of $Q^*$ on the duration $\t$ and the compression ratio $\g$ for a linear protocol $\o^2(t) =\o^2_i+(\o^2_f - \o^2_i) t/\t$, $0\leq t \leq \t$.

For the measurement stroke we consider a position  measurement of the oscillator. A projective measurement of position is not feasible because the eigenfunctions of the position operator do not possess a finite norm and hence do not belong to the Hilbert space of the oscillator. We rather apply a Gaussian window peaked at the target position $x$ and characterized by a finite width $\s$ \cite{FL}. This windowing operation transforms a state $\r$ prior to the measurement into the non-normalized post-measurement state $\r^{\text{pm}}_x$ given by
\be
\r^{\text{pm}}_x
=\frac{1}{\sqrt{2 \p \s^2}} e^{-(q-x)^2/(4 \s^2)} \r  e^{-(q-x)^2/(4 \s^2)}\:.
\ee{rpm}
The probability to find the target value $x$ follows as $p_x= \Tr \r^{\text{pm}}_x = \Tr e^{-(q-x)^2/(2 \s^2)} \r/\sqrt{2 \p \s^2} $. In the present situation we are not interested in the particular measurement outcome but only want to make use of the state transformation due to a non-selective measurement. This is given by the operation $\F^{\mathcal{M}}$ acting on the density matrix  immediately before the  measurement to yield the following post-measurement state: 
\be
\begin{split}
\F^{\mathcal{M}} (\r) & = \i_{-\infty}^\infty dx \r^{\text{pm}}_x\\
&= \i_{-\infty}^\infty \frac{dx}{\sqrt{2 \p \s^2}}   e^{-(q-x)^2/(4 \s^2)} \r  e^{-(q-x)^2/(4 \s^2)} \:,
\end{split}
\ee{FM}
where the integration over $x$ reflects the fact that the results of the measurement are ignored.
Note that the operation $\F^{\mathcal{M}}$ is unital and that hence it increases the average energy of the system provided the energy distribution before the measurement decreases with increasing energy, see Section~\ref{M}. 

Using the position representation of the eigenstates of an oscillator with the frequency $\o_f$ \cite{Schiff}
\be
\begin{split}
\ps_n(q) &\equiv \langle q | n; \o_f \rangle\\
& = \frac{1}{\sqrt{2^n n!}} \left (\frac{m \o_f}{\p \hbar} \right )^{1/4} e^{-m \o_f q^2/(2 \hbar)} H_n \left (\sqrt{\frac{m \o_f}{\hbar} }q \right )\:,
\end{split}
\ee{psn}
where $H_n(x)$ denotes the Hermite polynomial of order $n$,  
one obtains for the transition probability induced by the measurement the expression
\be
\begin{split}
T_{\mathcal{M}}(m,n)& =\i \frac{dx}{\sqrt{2 \p \s^2}} dq dq' \ps_n(q) \ps_m(q) \ps_n(q') \ps_m(q') \\
& \quad \times e^{-\left [ (q-x)^2 + (q'-x)^2 \right ] /(4 \s^2)}\:.
\end{split}
\ee{TMmn}
The $x$-integration can be performed exactly to yield
\be
\begin{split}
T_{\mathcal{M}}(m,n)& = \frac{1}{2^{n+m} n! m! \p}\\
&\quad \times \i d\z d\z' H_n(\z) H_m(\z) H_n(\z') H_m(\z')\\
&\quad \times e^{-\frac{1}{2}\bfz \cdot \bM^{-1} \cdot \bfz^T}\:,
\end{split}
\ee{TMmnM}  
where $\bfz$ is a vector with the components $\z$, $\z'$, and $\bM$ denotes a symmetric matrix of the form
\be
\bM = \left ( \begin{array}{cc}
\frac{8 r+1}{4(4 r +1)} & \frac{1}{4(4r +1)}\\
\frac{1}{4(4 r+1)} & \frac{8 r+1}{4(4 r+1)}
\end{array}
\right ) \:.
\ee{bM} 
The dimensionless parameter $r$ defined as
\be
r=\s^2 m \o_f /\hbar
\ee{r}
  equals twice the ratio of the  variances of the measurement window $\s^2$ and of the oscillator position in the ground-state $\hbar/(2 m \o_f)$. The larger $r^{-1}$ the more precise is the position measurement. A closed analytic expression of the double integral representing the transition probability in Eq.~(\ref{TMmnM}) is not known. Also a numerical integration is possible only for relatively small indices $n,m$ because of the pronounced oscillations of the integrand. Expressions for larger indices can be obtained by writing the double integral as a Gaussian average over four Hermite polynomials, for details see the Appendix~\ref{TMG}. The resulting expression can be summed up as long as $n,m \leq 12$. For larger values numerical problems prohibit reliable results. Alternatively, a closed expression for a generating function of the transition probability can be obtained, see the Appendix~\ref{TMgf}. Also this strategy is limited because with the increasing  order of the partial derivatives of the generating function required for the calculation of the transition probabilities the  available storage capacity of any computer will be reached at some point. The numerical results presented below are based on the measurement transition probability between the first 21 states of a harmonic oscillator obtained by the second, characteristic function based, method.  
\begin{figure}
\includegraphics[width=0.5\textwidth]{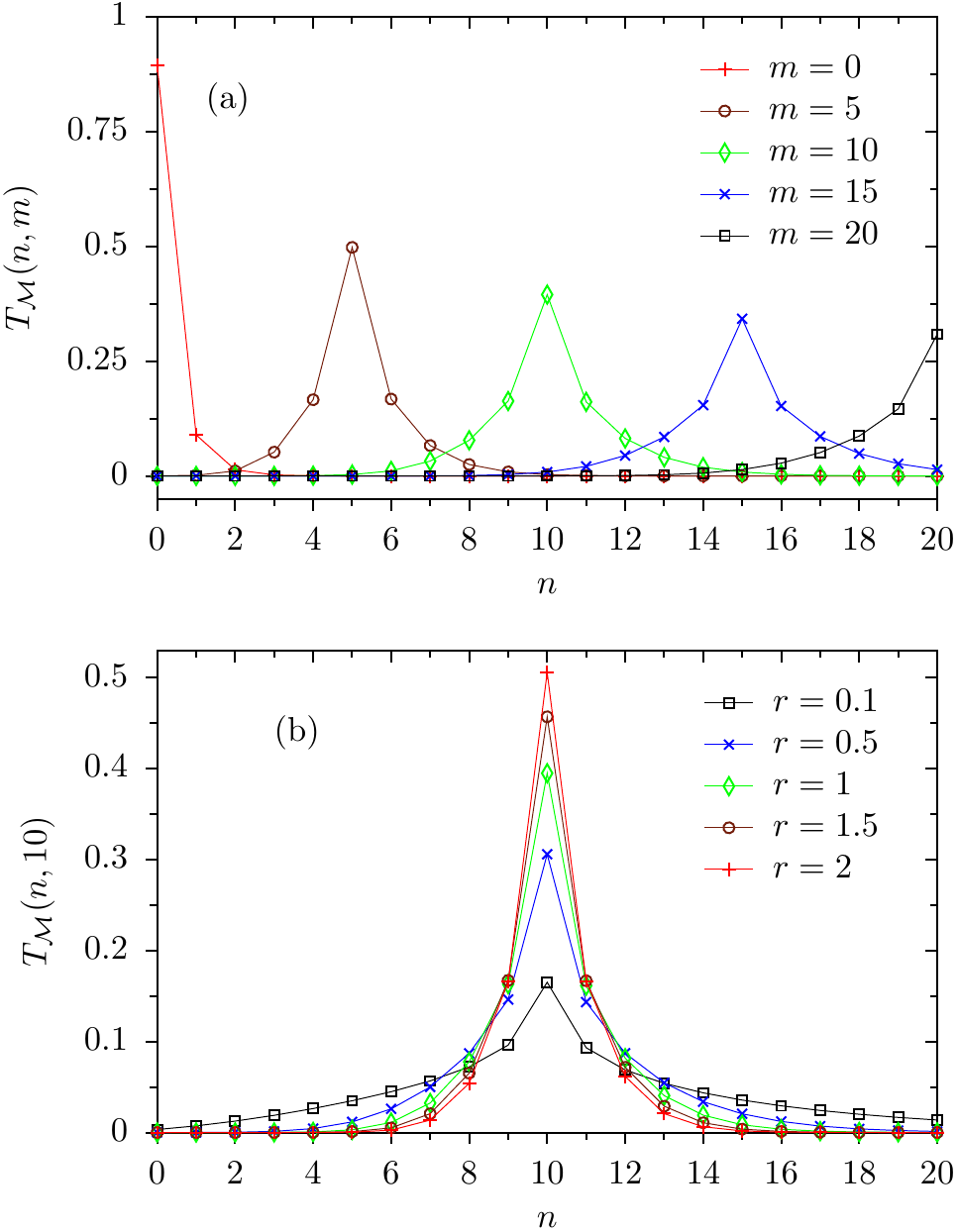}
\caption{The measurement-induced transition probability $T_{\mathcal{M}}(n,m)$ as a function of the target state $n$ is displayed in panel (a) at the fixed measurement parameter $r=1$ for different initial states, and in panel (b) for the initial state $m=10$ and different measurement parameters $r$. The initial state always has the largest probability, i.e. most likely, the measurement does not trigger a transition as can be seen from panel (a). With increasing precision of the position measurement (decreasing $r$) the transitions to more remote states become more likely. The probability for a transition decreases with the distance between the final and the initial  state $m-n$. Positive and negative distances with the same absolute value occur with slightly different probabilities, see also Fig.~\ref{bias} which displays the bias towards excitation over decay caused by a Gaussian position measurement. The thin lines serve as guides to the eye.}
\label{fTM}
\end{figure}  
\begin{figure}
\includegraphics[width=0.45\textwidth]{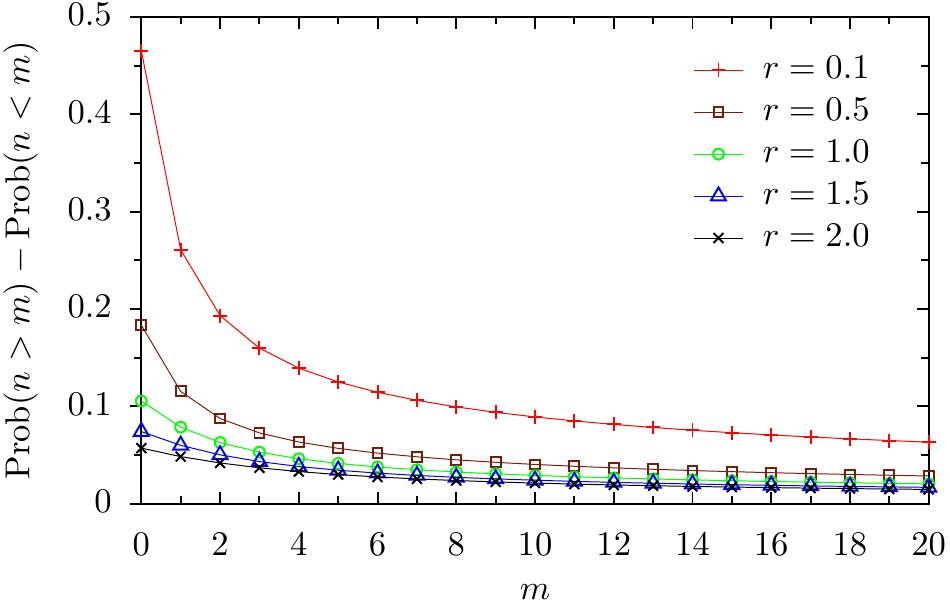}
\caption{The bias $\text{Prob}(n>m)-\text{Prob}(n<m) =\sum_{n>m} T_{\mathcal{M}}(n,m) - \sum_{n<m} T_{\mathcal{M}}(n,m)$ of measurement-induced transitions is displayed as a function of the initial state $m$ for different measurement parameters $r$. This bias is positive implying that more transitions go to higher than to lower excited states as seen from the initial state. For smaller parameters $r$ corresponding to more precise position measurements the bias becomes larger. It decreases with increasing excitation of the initial state and apparently approaches a constant positive value for sufficiently large values of $m$. The thin lines serve as a guide to the eye.}
\label{bias}
\end{figure}

The transition probabilities $T_{\mathcal{M}}(m,n)$ for a given initial state $n$ have a single maximum as a function of $m$ at $m =n$ as exemplified in Fig~\ref{fTM}. The probability $\text{Prob}(m>n) = \sum_{m>n} T_{\mathcal{M}}(m,n) $ characterizing   transitions from a fixed state $n$ to all states with $m>n$  is larger than the probability $\text{Prob}(m<n) = \sum_{m<n} T_{\mathcal{M}}(m,n) $ to loose energy. With increasing index $n$, the difference between these probabilities approaches a constant value, see Fig.~\ref{bias}. Moreover, as  demonstrated in Ref.~\cite{YTK}, the average value of the population difference between the state after and the one before a measurement is independent of the initial distribution of states. It is given by  
\be
\begin{split}
\hbar \o_f\left (\langle n \rangle_{\r^{\text{pm}}} - \langle n \rangle_\r \right )& =\Tr H(\o_f)\left (\F^{\mathcal{M}}(\r) -\r \right )\\
& = \hbar \o_f/(8 r)
\end{split}
\ee{mE}
and, hence, only depends on the strength of the measurement, see also the Appendix \ref{avdn}.     

While the ideal thermalization in the final stroke described by Eq. (\ref{Tb}) with  
\be
p_\b(n) =(1-e^{-\b \hbar \o_i}) e^{-\b \hbar \o_i n}
\ee{pboi}
takes an infinite amount of time, we shall also consider an imperfect thermalization of the working substance within a finite amount of time, $\th$, a process caused by the weak interaction of the working substance with a heat-bath at the required temperature. In the simplest way, this process can be modeled by a Markovian master equation of the form \cite{WH}          
\be
\begin{split}
\dot{\r}(t)& = - i \o_i [a^\dagger a, \r(t)]  + \g_\downarrow \left ( [a,\r(t) a^\dagger] +[a \r(t), a^\dagger] \right )\\
&\quad + \g_\uparrow \left ( [a^\dagger,\r(t) a] + [a^\dagger \r(t), a ] \right ) \:,
\end{split}
\ee{meWH}
where $a$ and $a^\dagger$ denote the annihilation and creation operators of the harmonic oscillator with frequency $\o_i$. The parameters $\g_\downarrow$ and $\g_\uparrow$ result from the interaction of the oscillator with the heat-bath; their ratio $\n \equiv \g_\uparrow/\g_\downarrow = e^{-\b \hbar \o_i}$ is given by the temperature of the heat-bath. Their difference $\k = \g_\downarrow -\g_\uparrow$ determines the relaxation rate of the mean value of the annihilation operator according to $\langle a(t) \rangle = e^{(i \o_i - \k )t} \langle a(0) \rangle$ for $t\geq 0$. The underlying assumption of weak coupling between the oscillator and the heat-bath implies that $\k \ll \o_i$ and also $\hbar \k \b \ll 1$, \cite{GWT}. Due to the rotating wave approximation inherent in the master equation (\ref{meWH}) the time evolution of the diagonal terms $p(n;t) = \Tr P_n(\o_i) \r(t) = \langle n;\o_i|\r |n;\o_i \rangle$ decouples from the non-diagonal terms of the master equation. The diagonal terms satisfy the following classical birth-death master equation   
\be
\begin{split}
\dot{p}(n;t) &= 2 n \g_\uparrow p(n-1;t) +2(n+1) \g_\downarrow p(n+1;t)\\
&\quad -2\left ( (n+1) \g_\uparrow +n \g_\downarrow \right ) p(n;t) \:.  
\end{split}
\ee{bdme}
Its solution is known in terms of conditional probabilities $p(n;t|l)$ satisfying the initial condition $p(n;0|l) = \d_{n,l}$ \cite{KMG,GRD}. Choosing for $t$ the thermalization time $\th$ one finds for the transition probability $T_\b(n,l) = p(n;\th|l)$ the expression
\begin{widetext}
\be
T_\b(n,l) = \frac{1-\n}{1-\l \a} \n^n \sum_{i=0}^{\min(n.l)} \frac{(-1)^i (n+l-i)!}{(n-i)! (l-i)! i!} \left( \frac{1-\a}{1-\n \a} \right )^{n+l-i} \left ( \frac{1- \a/ \n}{1-\a} \right )^i\:,
\ee{Tbnl} 
\end{widetext}
where $\a = e^{-2 \k \th}$ is a dimensionless parameter determining how far the thermalization has proceeded within the time span $\th$. A quantitative measure of the degree of thermalization is provided by the $\ell_1$-norm of the difference between the actual probability distribution $p(t)=(p(n;t), n=0,1, \ldots)$ and the target distribution $p_\b=(p_\b(n), n=0,1,\ldots)$, 
\be
\begin{split}
d(p(t),p_\b) &= ||p(t) -p_\b ||_1\\
& = \sum_{n=0}^\infty |p(n;t) -p_\b(n)|\:.  
\end{split}
\ee{1norm}
\section{Performance of the harmonic oscillator measurement engine}\label{perf}
In the case of perfect thermalization, the above described engine is characterized by the four dimensionless parameters, $\hbar \b \o_i$ specifying the inverse temperature of the initial state which is that of the heat-bath, the duration of the two work strokes $\o_i \t$,  the compression rate $\g=\o_f/\o_i$ and the ratio $r=\s^2 m \o_f/\hbar$ determining the precision of the position measurement. The two work stroke parameters $\o_i \t$ and $\g$ determine the Husimi parameter $Q^*$, which additionally depends on the particular protocol specifying how the frequency interpolates between the initial and the final values. Here we will always assume that $\o^2(t) = (\o^2_f -\o^2_i) t /\tau + \o^2_i$ follows a linear protocol.
For imperfect thermalization, with the damping strength $\k$ and the duration of the thermalization stroke $\th$, two further dimensionless parameters $\k/\o_i$ and $\o_i \th$, enter. In order that the Markovian master equation provides a valid modeling, as already mentioned, the damping parameter must be small compared to the oscillator frequency and the Matsubara frequency, i.e. $\k  \ll \o_i, k_B T/\hbar$. To achieve a well thermalized final state the total time must be sufficiently large, $\o_i \th \gg 1$ such that $\k \th > 1$. 
\subsection{The statistics of work and supplied energy}
The total work $W$ is given by the sum of works $W_I$ and $W_{II}$
\be
W =W_I + W_{II}=E_1 -E_0+E_3 -E_2\:,    
\ee{W}
according to Eq. (\ref{WI}) and Eq. (\ref{WII}).
Therefore, the joint probability density function (pdf) $\r(W,Q)$ of the total work $W$ and the measurement supplied energy $E_\mathcal{M}$ defined in Eq. (\ref{EM}) is given by 
\be
\begin{split}
\r(W,E_{\mathcal{M}})& =\sum_{\substack{ k,l,m,n\\ =0}}^\infty \d(W-\hbar \o_f(m-k) -\hbar \o_i(l-n))\\
&\quad \times \d(E_{\mathcal{M}} -\hbar \o_f(k-m)) p^{(4)}(l,k,m,n)\\
&= (\hbar \o_i)^{-2} \sum_{\substack{w,z\\=-\infty}}^\infty  \d(W(\hbar\o_i)^{-1}-w+ \g z) \\ 
& \quad \times \d(E_{\mathcal{M}}(\hbar \o_i)^{-1} - \g z) p_{w,z}(w,z)\:,
\end{split}
\ee{rwe}  
where the 4-point probability $p^{(4)} (l,k,m,n) = \sum_{n'} p(n',l,k,m,n)$ follows from Eq. (\ref{p5}) as
\be
p^{(4)} (l,k,m,n) =
T(k,l) T_{\mathcal{M}}(k,m)T(m,n) p_\b(n)\:.
\ee{plkmn} 
The two-point probabilities $p_{w,z}(w,z)$ of finding $w= l-n$ and $z=k-m$ which specify the population differences of the states {\bf 3} and {\bf 0}, and of {\bf 2} and {\bf 1}, respectively, are given by  
\be
p_{wz}(w,z) = \sum_{m,n} T(z+m,z+n) T_{\mathcal{M}} (z+m,m) T(m,n) p_\b(n)\:.
\ee{pwz}

In the particular case of adiabatically slow work strokes, i.e. in the limit $\t \to \infty$, one finds with $T^{\text{ad}}(k,l) = \d_{k,l}$ for the joint probability
\be
\r^{\text{ad}}(W,E_{\mathcal{M}}) = \delta(W +(1-\g^{-1}) E_{\mathcal{M}}) \r^{\text{ad}} (E_{\mathcal{M}})\:,
\ee{rad}
where 
\be
\r^{\text{ad}}(E_{\mathcal{M}}) = \sum_{n,l} \d(E_{\mathcal{M}}- \hbar \o_f (l-n)) T_{\mathcal{M}}(l,n) p_\b(n)
\ee{rEad} 
determines the pdf of the measurement supplied energy. Hence, the random work and supplied energy are strictly related to each other as $W= -(1-\g^{-1} ) E_{\mathcal{M}}$. The resulting efficiency $\h = -W/E_{\mathcal{M}} = 1-\g^{-1}$ agrees with Eq.~(\ref{hg}).

In the general, non-adiabatic, case, the statistics of total work and supplied energy is determined by the marginal distribution of the supplied energy $E_{\mathcal{M}}$ 
\be
\r_{\mathcal{M}}(E_{\mathcal{M}}) = \sum_{z=-\infty}^\infty \d(E_{\mathcal{M}} -\hbar \o_i \g z) p(z)
\ee{rM}
and by the conditional probability
\be
\r(W|E_{\mathcal{M}}\!=\!\hbar \o_i \g z) = \!\sum_{w=-\infty}^\infty\! \d (W-\hbar \o_i( w - \g z)) p(w|z)\:,
\ee{crWE}
where
\be
p(z)= \sum_{m,n}
T_{\mathcal{M}}(m+z,m) T(m,n) p_\b(n)
\ee{pz}
and
\be
p(w|z) = \frac{p_{wz}(w,z)}{p(z)} \:.
\ee{cpwz}
\begin{figure}
\includegraphics[width=0.45\textwidth]{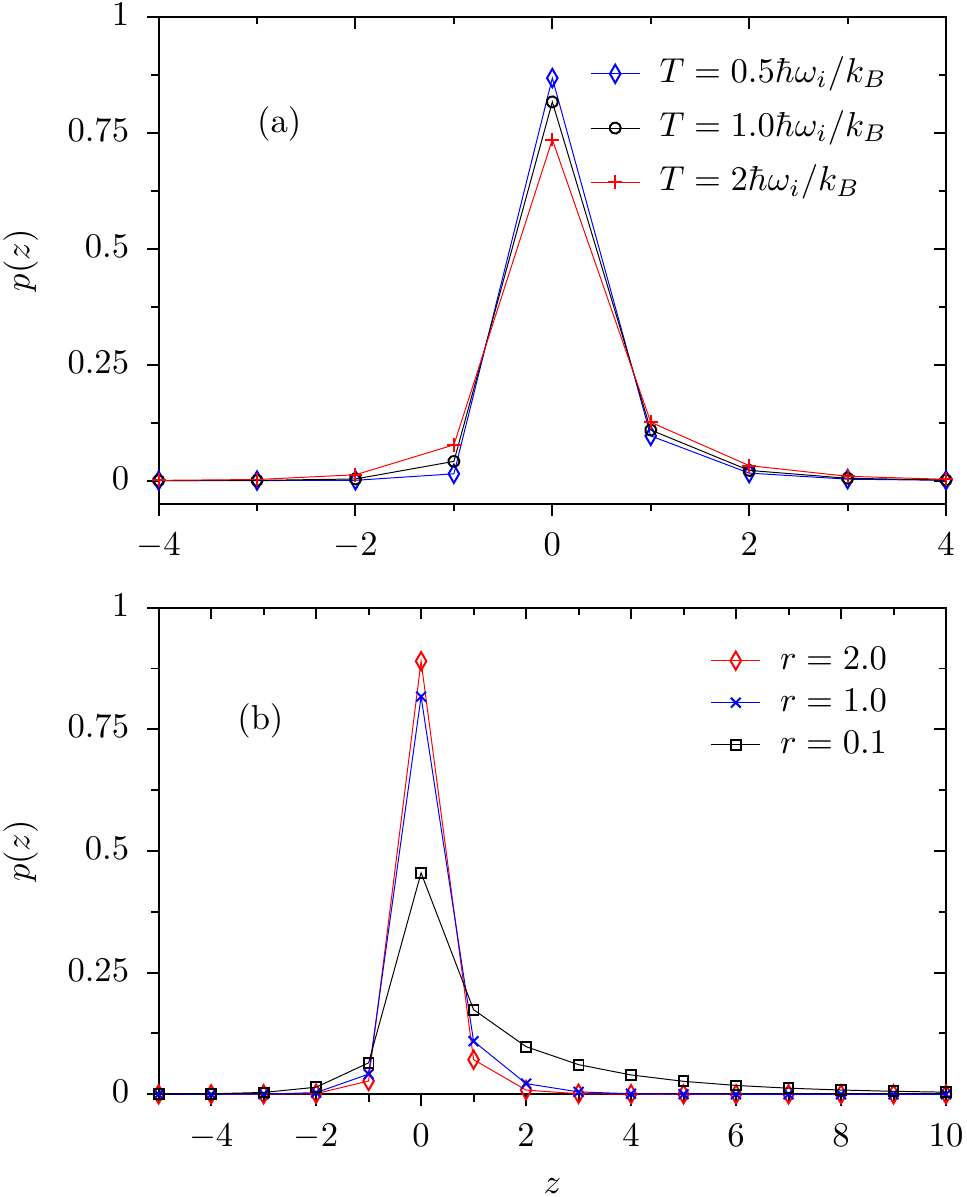}
\caption{The probability $p(z)$ defined in Eq. (\ref{pz}) of a population difference of size $z$ between the states {\bf 2} and {\bf 1} is displayed for different parameter values of the engine. The difference $z$ determines the random energy input $E_{\mathcal{M}} = \hbar \o_f z$. In all cases, it is most likely that the measurement does not cause a transition and most of the time no energy is injected in the measurement stroke. Yet, all distributions are biased towards a positive value of $z$ and hence to an increase of energy. Panel (a) exemplifies the influence of the initial temperature $T$  at the measurement parameter $r=1$ and Husimi parameter $Q^*=1.05$. Increasing temperature leads to a wider distribution of $z$. In panel (b) the  temperature is kept fixed at $k_B T =\hbar \o_i$ as well as the Husimi parameter at $Q^* = 1.05$ for different measurement parameters $r$. A decreasing measurement parameter leads to a broadening while a variation of the Husimi parameter does not visibly influence the $z$-distribution. Apart from the dependence on the Husimi parameter there is no direct dependence of $p(z)$ on the compression ratio $\g$. The thin lines serve as a guide to the eye.
       }
\label{PZ}
\end{figure}
\begin{figure}
\includegraphics[width=0.5\textwidth]{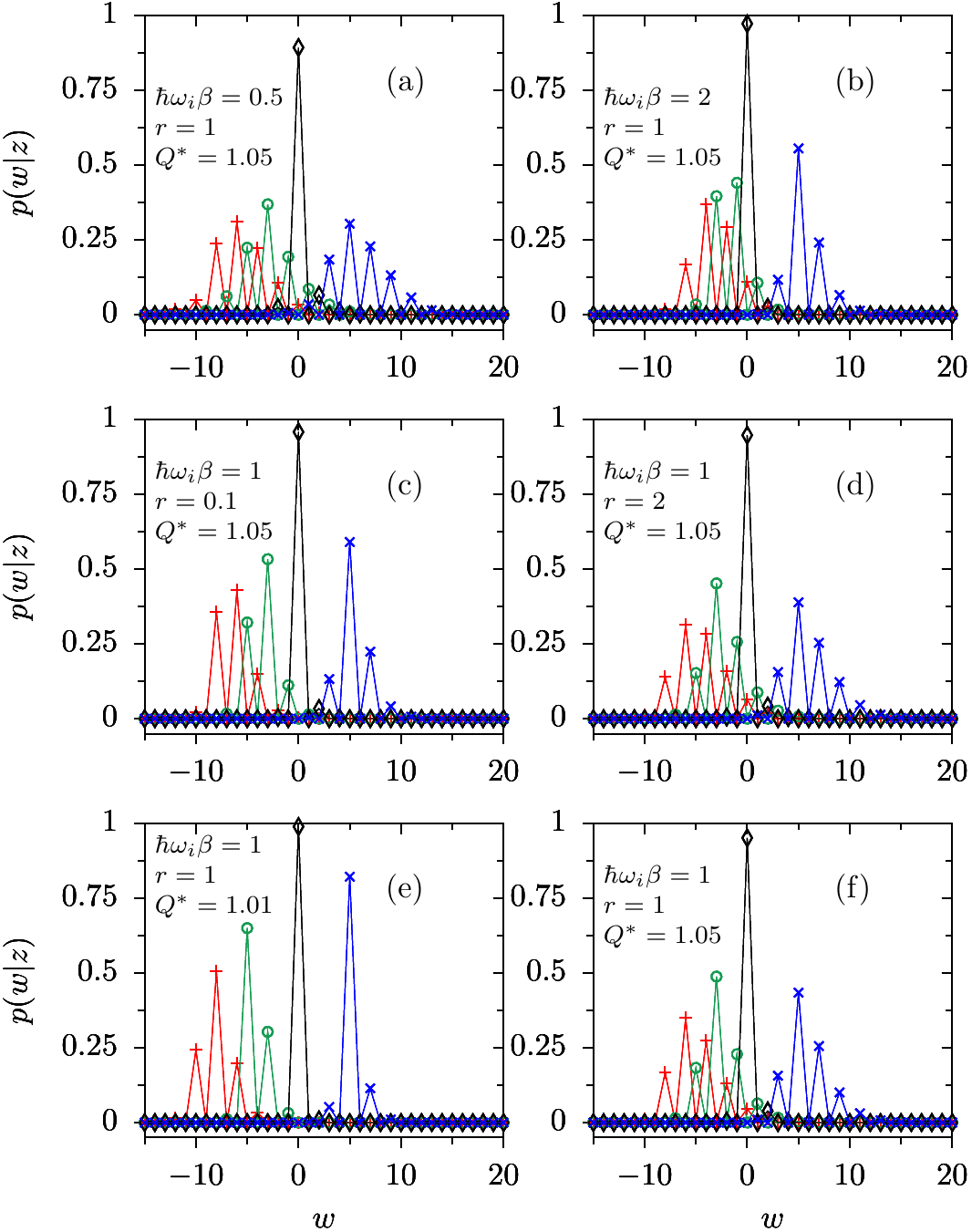}
\caption{The probabilities $p(w|z)$, see Eq. (\ref{cpwz}) are displayed for the different conditions $z=-10$ (red pluses), $-5$ (green circles), $0$ (black diamonds) and $5$ (blue crosses)   at different parameter values as indicated in each panel. In each row one of the parameters $\b$, $r$ and $Q^*$ is varied while the other two are kept constant.
Panel (a) and (b) confirm a narrowing of the distributions when the temperature is decreased; similarly a larger measurement parameter leads to a narrower distribution whereby the effect is only modest for a variation of $r$  by a factor of 20 in (c) and (d). Likewise, the increase of the Husimi parameter entails a broadening of the distribution as exemplified in panels (e) and (f). The conditional probability depends only via the Husimi parameter $Q^*$ on the compression ratio $\g$. Lines are meant as a guide to the eye.} 
\label{CPWZ}
\end{figure}
The Figs.~\ref{PZ} and \ref{CPWZ} exemplify the two auxiliary distributions $p(z)$ and $p(w|z)$, respectively. In the adiabatic limit the conditional probability simplifies to a Kronecker delta: $p^{\text{ad}}(w|z) = \d_{w,z}$. Because $p(z)$ is maximal at $z=0$ the most probable value of $E_{\mathcal{M}}$ vanishes. Yet, the distribution $p(z)$ is slightly biased 
towards positive values leading to a positive energy supply on average, in accordance with the findings of Ref.~\cite{YTK}, see also the Appendix~\ref{avdn}.
\begin{figure}
\includegraphics[width=0.45\textwidth]{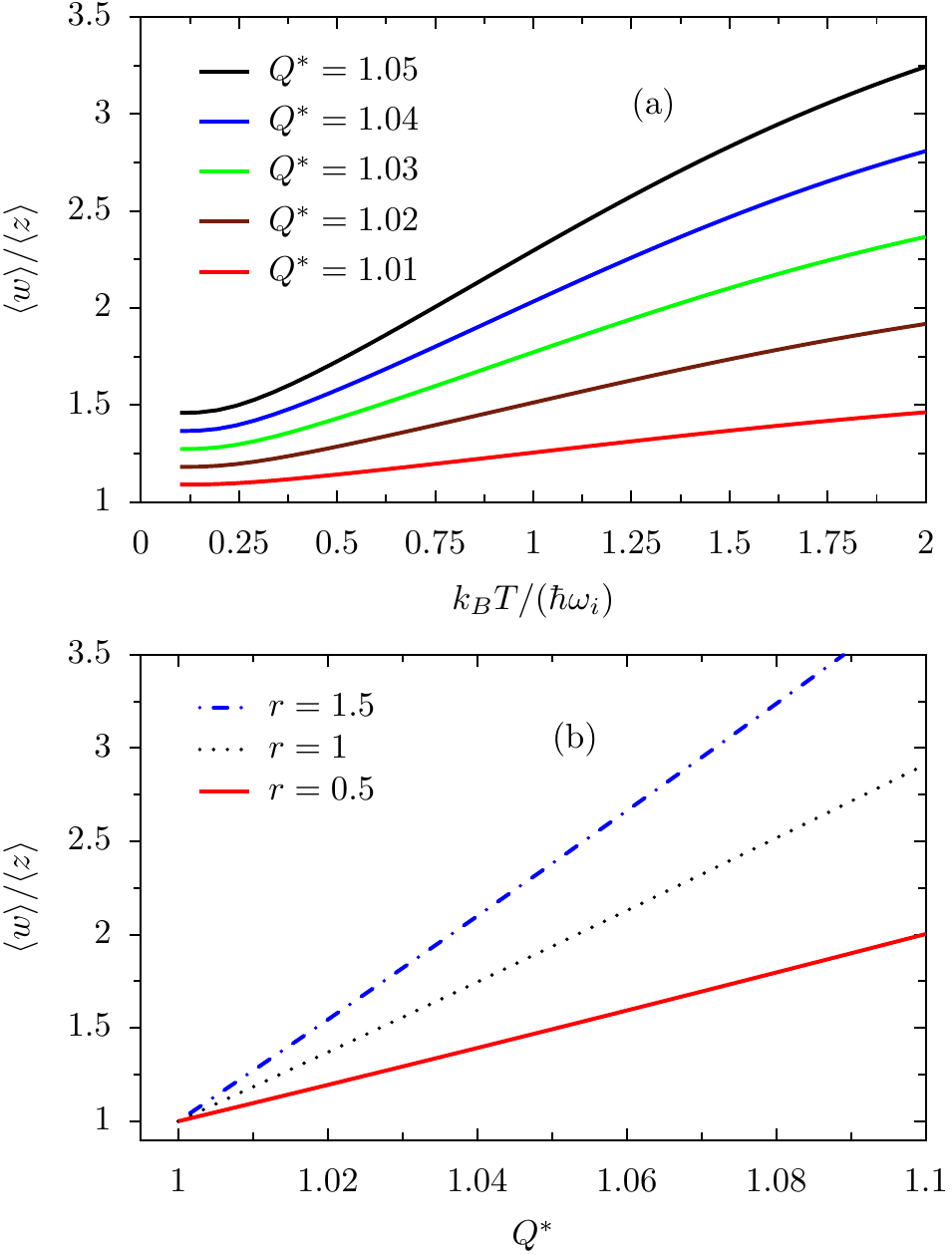}
\caption{The ratio $\langle w\rangle/ \langle z \rangle$ specifying the relation of the average population difference between the states {\bf 3} and {\bf 0} and that between the states {\bf 2} and {\bf 1}, respectively, is displayed as a function of temperature for increasing values of the Husimi parameters $Q^*$ in panel (a) from bottom to top at the fixed measurement parameter $r=1$. The curves lie above each other according to the increasing magnitude of $Q^*$. In panel (b) the dependence of $\langle w\rangle/ \langle z \rangle$ on the Husimi parameter is exemplified for different measurement parameters $r$ at the fixed temperature $k_B T = \hbar \omega_i$. According to Eq. (\ref{h}) the ratio  $\langle w\rangle/ \langle z \rangle$ agrees with the value of the minimal compression coefficient $\g$. At smaller compression coefficients the efficiency becomes negative and the device no longer acts as an engine by consuming energy without doing work. Note that the ratio $\langle w\rangle/ \langle z \rangle$ does depend on $Q^*$ but has no explicit dependence on the compression ratio $\g$.   }
\label{woz}
\end{figure}
\begin{figure}{t}
\includegraphics[width=0.45\textwidth]{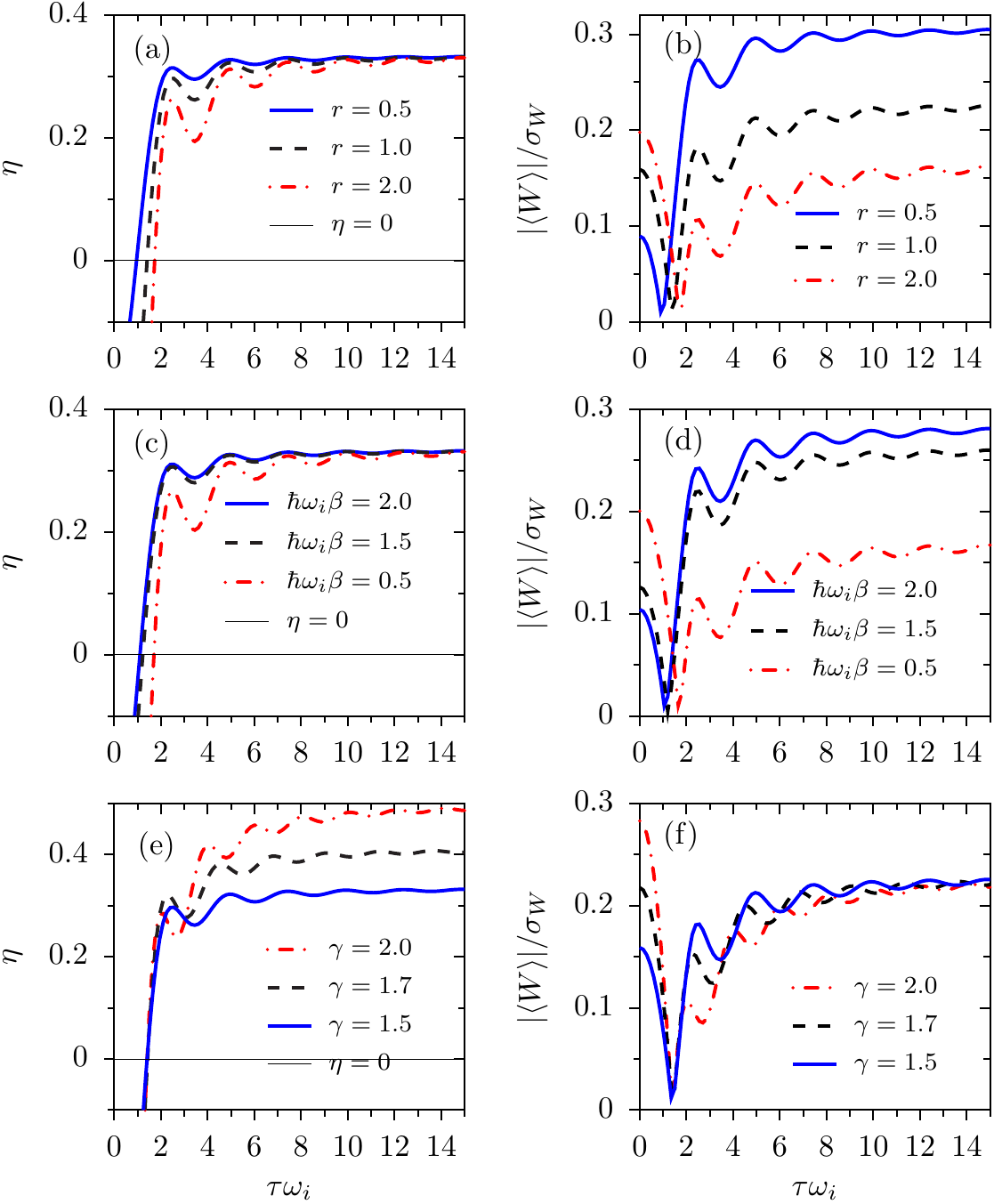}
\caption{The performance of the engine is illustrated as a function of the work stroke time $\t$ in terms of the efficiency $\h$ (left column) and the reliability $|\langle W \rangle| / \s_W = (\langle W^2 \rangle/ \langle W \rangle^2 -1)^{-1/2}$ (right column). In the first row the efficiency (a) and the reliability (b) are displayed for different measurement parameters $r$ at the inverse temperature $\b \hbar \o_i=1$ and compression rate $\g=1.5$. Accordingly, for large values of $\t$ the efficiencies converge to the same adiabatic value $\h^{\text{ad}} = 1-\g^{-1} =1/3$, see Eq. (\ref{hg}). For small $\t$, the efficiency is negative indicating a malfunctioning engine. The work stroke time at which the efficiency becomes positive increases with increasing measurement parameter. For all stroke times with positive efficiency, the reliability  increases with decreasing measurement parameter. 
Panels (c) and (d) expose the impact of different temperatures $T$ at fixed $r=1$ and $\g=1.5$. While the influence of temperature on the efficiency is relatively week, an increase of temperature  leads to a marked worsening of the reliability. The bottom row illustrates the effect of the compression rate  $\g $. While the efficiencies approach different adiabatic limits in accordance with Eq. (\ref{hg}), the reliabilities converge to $\g$-independent asymptotic values given by E. (\ref{wv}).   Panels (e) and (f) exemplify the dependence on the compression rate at fixed $\b \hbar \o_i =1$ and $r=1$. 
In all cases, the efficiencies and the reliabilities display oscillations in dependence of the stroke time $\t$, similarly as the Husimi parameter, see also Fig.~\ref{hus}.
}       
\label{RELY}
\end{figure}

Both the work and the supplied energy can be written in terms of the integer random variables $w$ and $z$ as $W/(\hbar \o_i) = w - \g z$ and $E_{\mathcal{M}}/(\hbar \o_i )= \g z$. Due to the linearity of these relations the average work and supplied energy can be expressed by the averages of $w$ and $z$ as
\begin{align}
\langle W \rangle /(\hbar \o_i) &= \langle w \rangle - \g \langle z \rangle\:, \label{AvW}\\
\langle E_{\mathcal{M}} \rangle/(\hbar \o_i) &= \g \langle z \rangle \:.
\label{AvE}
\end{align}
As a consequence, the efficiency becomes 
\be
\h = - \frac{\langle W \rangle}{\langle E_{\mathcal{M}} \rangle} = 1 - \g^{-1} \frac{\langle w \rangle}{\langle z \rangle} \:.
\ee{h}
For expressions of the average values $\langle w \rangle$ and $\langle z \rangle$ in terms of the stroke transition probabilities see the Appendix~\ref{Awe}. 

In the adiabatic case the averages of $w$ and $z$ coincide and the known result (\ref{hg}) is recovered. In general, the ratio $\langle w \rangle/\langle z \rangle$ is larger than one. It is a function of the Husimi parameter and temperature, see Fig.~\ref{woz}. In particular at too rapid work strokes and also at too high temperatures of the initial state, this ratio may become larger than the compression rate with the result that, with negative  efficiency and positive total work, the device no longer functions as an engine.  

As a result of the relatively weak positive bias of the probability distribution of $z$ towards positive values, the root mean square displacement of the supplied energy excels  the average value considerably, being smallest for adiabatic work strokes. Accordingly, also the average work is smaller than the root mean square deviations of the total work, as displayed in Fig.~\ref{RELY}. The efficiency given by the ratio of the average negative total work and the average  supplied energy has its maximal value for adiabatic work processes and decreases with shorter work stroke times as exemplified in 
the left column of Fig.~\ref{RELY}. Eq. (\ref{h}) in combination with the approximately linear behavior of the ratio $\langle w \rangle /\langle z \rangle$ on the Husimi parameter $Q^*$, see Fig.~\ref{woz}(b), leads to a roughly linear decrease of the efficiency as a function of the Husimi parameter until, as already mentioned, for too fast work strokes a critical value of the Husimi parameter $Q^{*}$ parameter is reached beyond which the efficiency becomes negative; on average, the engine then dissipates more energy than it receives by the measurement stroke.

For adiabatic work strokes the second moment of the energy supplied to the harmonic oscillator can be analytically determined. By writing this moment as 
\be
\langle E^2_{\mathcal{M}} \rangle = (\hbar \o_f)^2 \sum_{n,l} (l-n)^2 T_{\mathcal{M}}(l,n) p_\b(n)
\ee{EM2}
one obtains with the expression (\ref{TMmn}) for the transition matrix $T_{\mathcal{M}}(l,n)$ after some tedious algebra the result
\be
 \langle E^2_{\mathcal{M}} \rangle  = \frac{(\hbar \o_i \g)^2}{8r} \left ( \frac{3}{8r} + \coth \frac{\hbar \b \o_i}{2} \right )\:.
\ee{E2}
The variance of the supplied energy characterizing its fluctuations becomes
\be
\begin{split}
\s^2_{\mathcal{M}} &= \langle E^2_{\mathcal{M}} \rangle - \langle E_{\mathcal{M}} \rangle^2 \\
&=\frac{(\hbar \o_i \g)^2}{8 r} \left ( \frac{1}{4r} + \coth \frac{\hbar \b \o_i}{2} \right )\:,
\end{split}
\ee{s2M}
signifying increasing fluctuations with increasing temperature $1/\b$, compression rate $\g$  and measurement strength $1/r$. Due to the rigid relation between supplied energy and performed work for adiabatic work strokes, see Eq. (\ref{rad}), one obtains for the variance of the work 
\be
\begin{split}
\s^2_W &= \langle W^2 \rangle - \langle W \rangle^2 \\
&=(\g^{-1}-1)^2 \s^2_{\mathcal{M}}\:.
\end{split}
\ee{s2W} 
Accordingly the covariance of the work and the 
supplied energy becomes
\be
\begin{split}
\s_{W,E_{\mathcal{M}}} &= \langle W E_{\mathcal{M}} \rangle - \langle W \rangle\langle E_{\mathcal{M}} \rangle \\
&=(\g^{-1}-1) \s^2_{\mathcal{M}}\:.
\end{split}
\ee{s2W} 
As a measure of reliability, the ratio of average work ($\langle W \rangle =(\g-1) /(8r))$) and the root mean spare deviation of the work follows from Eq. (\ref{s2W}) in the adiabatic limit as
\be
\frac{|\langle W \rangle|}{\s_W} = \left [2 + 8 r \coth \frac{\hbar \b \o_i}{2}  \right]^{-1/2}
\ee{wv}
which is independent of the compression ratio $\g$, as illustrated in the panel (f) of Figure~\ref{RELY}  by the convergence of the of the different reliability measures for slow work strokes, i.e. for large times $\t$.

\subsection{Imperfect thermalization}
\begin{figure}
\includegraphics[width=0.45\textwidth]{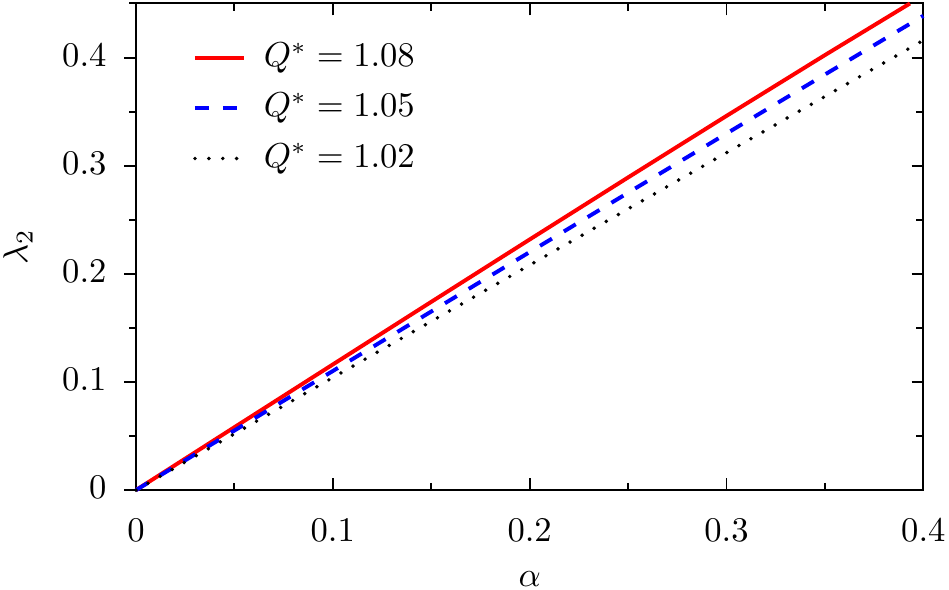}
\caption{The second largest absolute value of the eigenvalues $\l_2$ of the transition matrix $T_{\text{cycle}}$ defined in Eq. (\ref{Tc}) is displayed as a function of the thermalization parameter $\a = e^{- 2 \kappa \th}$ for different Husimi parameters $Q^* =1.02$ (black dash-dotted line), $1.05$ (blue dashed line) and $1.08$ (red solid line) at the temperature $T = \hbar \o_i$ and measurement parameter $r=1$. The graphs are indistinguishable from strait lines and vary only insignificantly with temperature and measurement parameter $r$. The transition matrix depends on the compression ratio $\g$ only via the Husimi parameter $Q^*$ as consequently does its second eigenvalue.}
\label{La2}
\end{figure}

\begin{figure}
\includegraphics[width=0.5\textwidth]{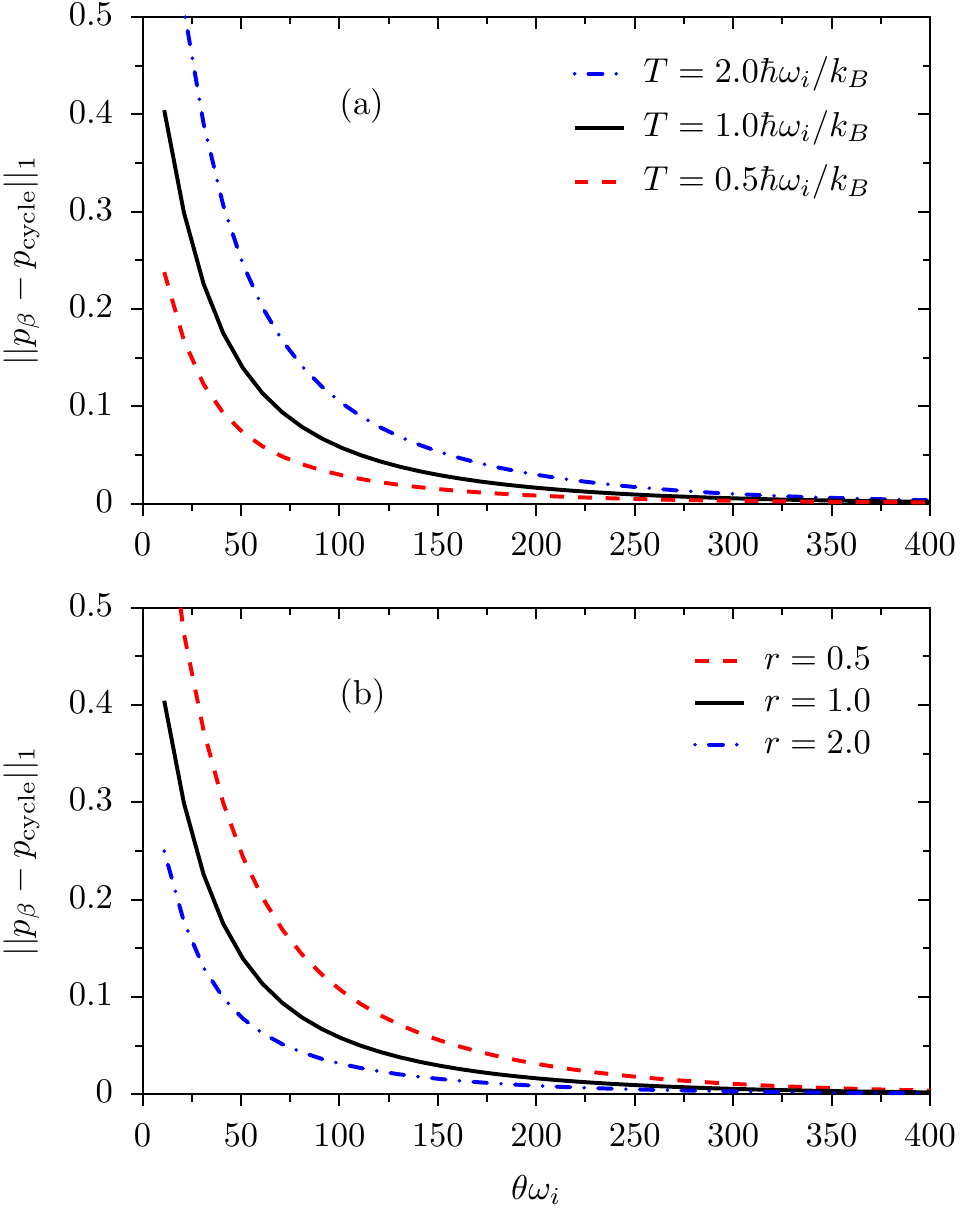}
\caption{The distance between the target state $p_\b$ and the state $p_{\text{cycle}}$ after a complete cycle at stationary operation is measured in terms of the 1-norm, Eq. (\ref{1norm}), and displayed as a function of the thermalization time $\th$ for different temperatures $T$ in panel (a) at the measurement parameters $r=1$. A smaller measurement parameter, $r=0.5$ (red dashed line) leads to a larger distance from the target state while the larger $r=2$ (blue dash dotted line) lessens this distance  as exemplified in panel (b) where the temperature is fixed at $T =\hbar \o_i/k_\b$. The  parameters specifying the work stroke only insignificantly influence the norm distance. With $\g=1.5$, and $\o_i \t =4 $ they are the same for all curves. For the relaxation rate of the thermalization always the same value $\k = 0.005 \o_i$ was chosen. 
}
\label{NORM}
\end{figure}

According  to Eq.(\ref{p5}), the probability to find the state $n'$ after a full cycle  when having started in the state $n$ is given by 
\be
T_{\text{cycle}}(n',n) = \sum_{l,k,m} T_\b(n',l)T(k,l) T_{\mathcal{M}}(k,m)T(m,n) \:.
\ee{Tc}     
If the final stroke of a cycle consists in a perfect thermalization, i.e. if $T_\b(n',l) = p_\b(n')$, also the transition matrix for a complete cycle becomes independent of the initial state, yielding 
\be
T^{\text{pt}}_{\text{cycle}} (n',n) = p_\b(n') \:.
\ee{Tpc}       
We model the impact of imperfect thermalization in terms of the weak coupling expression (\ref{Tbnl}). Because of the imperfect thermalization the final state differs more or less from the target Gibbs state depending on how the engine has started. This may lead to a transient behavior during a number of cycles until a stationary state is reached. Because the transition matrix $T_{\text{cycle}}$ is irreducible, the stationary state is independent of the initial one. It is given by the properly normalized eigenvector belonging to the eigenvalue $1$. The speed of the approach to the target Gibbs state can be estimated by the absolute value of the second largest eigenvalue of $T_{\text{cycle}}$ which, in the parameter range that we have investigated, is apparently proportional to the convergence parameter $\a =e^{-2 \k \th}$ characterizing the transition probability $T_\b(n',n)$ (Eq. (\ref{Tbnl})), see also Fig.~\ref{La2}. The approach to equilibrium is visualized in Fig.~\ref{NORM} as a function of the time of contact between the working substance and the heat bath in terms of the 1-norm, eq. (\ref{1norm})  of the difference between the actual and the target distribution.  


\begin{figure}
\includegraphics[width=0.5\textwidth]{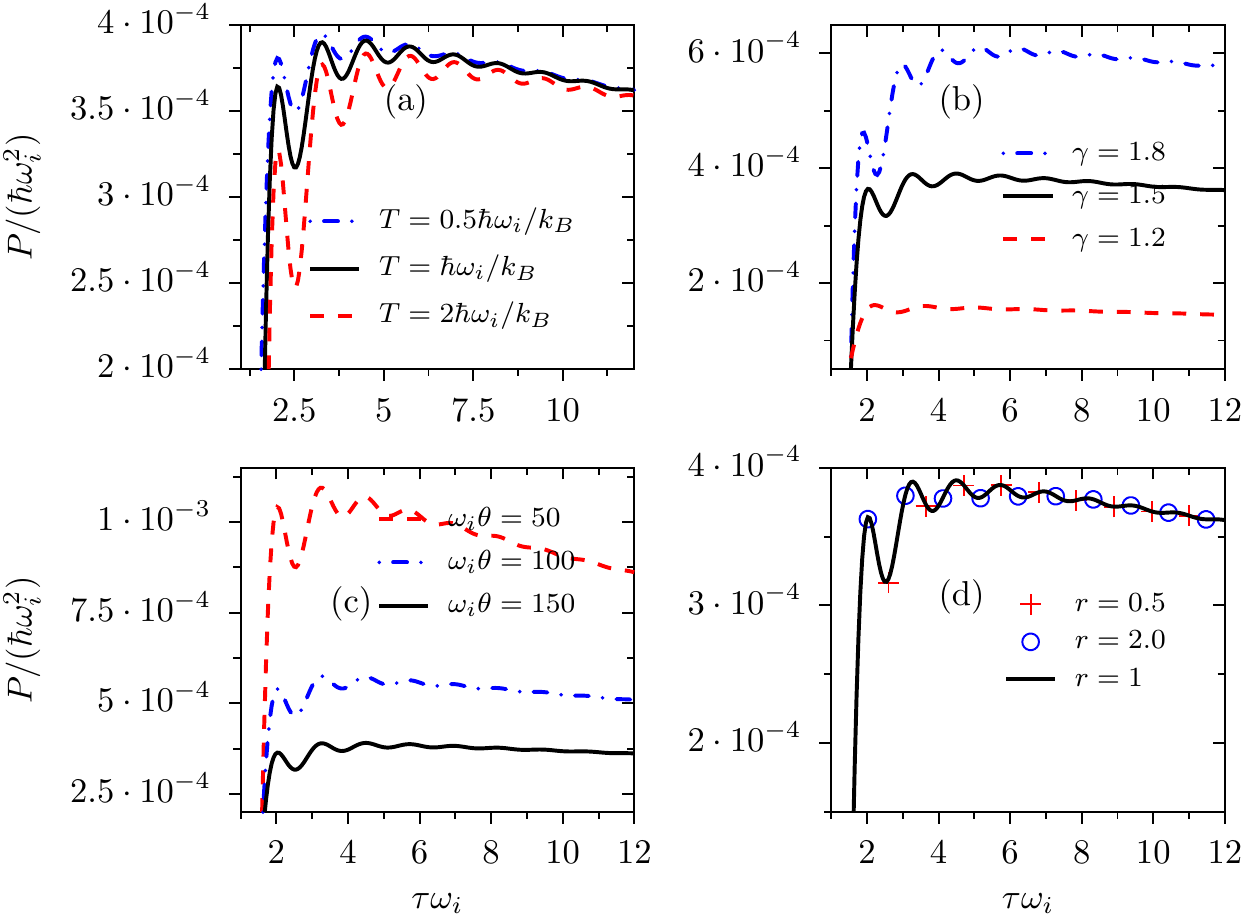}
\caption{The power of a measurement driven engine is displayed as a function of the duration of a work stroke $\t$, for a standard set of parameters $k_B T =\hbar \o_i$, $r=1$, $\g =1.5$ and $\th=150$ (black solid lines). In panel (a) its behavior is compared with the two temperatures $k_BT=2 \hbar \o_i $ (red dashed line), and $0.5 \hbar \o_i$ (blue dash dotted line) with the same other parameters. Likewise, the panel (b) illustrates the dependence on the compression rate with $\g = 1.2$ (red dashed line) and $\g=1.8$, in panel (c) the thermalization time is varied: $\th =50$ (red dashed line) and $100$ (blue dash dotted line), the variation of the measurement strength in panel (d) with $r=0.5$ (red pluses) and $2$ (blue circles) has only an insignificant influence on the power. The relaxation rate was always chosen as $ \k = 0.005 \o_i$.   }  
\label{POW}
\end{figure}

A complete cycle of an engine operated with non-adiabatic work strokes and imperfect thermalization takes a total time $t_{\text{cycle}} = 2 \t + \th$, where each work stroke lasts the time  $\t$ and the thermalization stroke takes the time $\th$. Now one can characterize the performance of the engine by the power $P = -\langle W \rangle/t_{\text{cycle}}$. In Fig.~\ref{POW} the average power is displayed as a function of the work stroke time $\t$ for different sets of the engine parameter values. At short times $\t$ the power is negative followed by a rapid increase to positive values in a similar way as the efficiency does. At intermediate work stroke times it assumes its largest values superimposed by oscillations. Finally the power slowly decays to zero in the adiabatic limit. The oscillatory behavior is caused by the according dependence of the Husimi parameter $Q^*$ on the work stroke duration, see also Fig.~\ref{hus}.

\section{Comparison with a two-temperature Otto engine}\label{Otto}
As already mentioned, the present device differs from a more conventional Otto engine only by the stroke from state {\bf 1} to state {\bf 2} with which energy is supplied by the coupling to a hot heat bath rather than to a measurement apparatus. For the sake of simplicity we shall assume that this stroke leads to a complete thermalization at the inverse hot temperature $\b_h$. Also the final stroke is supposed to perfectly thermalize at the inverse low temperature $\b_c$. Moreover we restrict the comparison to adiabatic work strokes. For this case, the joint probability $p^{\text{Otto}}(n',k,l,m,n)$ of finding within a complete cycle the states {\bf 0}, {\bf 1}, {\bf 2}, {\bf 3}, {\bf 0'} being excited at the levels $n$, $m$, $l$, $k$, $n'$ respectively,    is given by the expression
\be
p^{\text{Otto}}(n',k,l,m,n) = \d_{k,l} \d_{m,n} p^{(i)}_{\b_c}(n') p^{(f)}_{\b_h}(l) p^{(i)}_{\b_c}(n)
\ee{potto}
as it immediately follows from Eq. (\ref{p5}) with $T_{\mathcal{M}} (l,m)$ substituted by $T_{\b_h}(l,m) =p^{(f)}_{\b_h}(l)$, Eq. (\ref{Tb}), and from the transition probability (\ref{Tad}) for the adiabatic strokes. Here, the probabilities $p^{(\n)}_{\b_j}(n)= e^{-\b_j \hbar \o_\n n} (1- e^{- \b_j \hbar \o_\n})$ specify the  equilibrium distributions of an oscillator with frequency $\o_\n$, $\n =i,f$ at the temperature $\b_j$, $j=h,c$.  
For the measurement engine the population of state ${\bf 2}$ depends on that of the previous state ${\bf 1}$. In contrast for an Otto engine with perfect thermalization these populations are independent of each other. 
For the average and the second moment of the work one obtains
\begin{align}
\label{WO}
\langle W \rangle^{\text{Otto}}& = (1-\g)\hbar \o_i \left (\langle n \rangle^{(f)}_{\b_h} - \langle n \rangle^{(i)}_{\b_c} \right )\:,\\
\label{WO2}
\langle W^2 \rangle^{\text{Otto}}& = (1-\g)^2 (\hbar \o_i)^2 \nonumber\\
&\quad \times \left (\langle n^2 \rangle^{(f)}_{\b_h} -2 \langle n \rangle^{(f)}_{\b_h} \langle n \rangle^{(i)}_{\b_c} +\langle n^2 \rangle^{(i)}_{\b_c} \right )
\end{align}
yielding for the work variance of an adiabatically operated Otto-engine the
expression
\be
\s^2_{\text{Otto}}= (1-\g)^2 (\hbar \o_i) \left ( \s^2_n(\b_h,\o_f) +\s^2_n(\b_c,\o_i) \right )\:,
\ee{s2n}
where the first two moments of the thermal occupation numbers and the respective variance are known as
\begin{align} \label{n}
\langle n \rangle^{(\n)}_{\b_j} &= \frac{1}{e^{\b_j \hbar \o_\n}-1}\:,\\  
 \label{n2} 
\langle n^2 \rangle^{(\n)}_{\b_j} &= \frac{e^{\b_j\hbar \o_\n} +1}{(e^{\b_j\hbar \o_\n }-1)^2} \:,\\ \label{s2n}
\s^2_n(\b,\o)& = e^{\b \hbar \o}/(e^{\b \hbar \o}-1)^2\:,
\end{align}
with $\n$ standing for $i$ or $f$ and $j$ for $h$ or $c$.

Likewise, one finds for the average energy $\langle E_{\b_h}\rangle$ supplied by the contact with the hot heat bath the expression
\be
\langle E_{\b_h} \rangle = \hbar \o_f \left (\langle n \rangle^{(f)}_{\b_h} -\langle n \rangle^{(i)}_{\b_c} \right ) 
\ee{Ebh}
\begin{figure}
\includegraphics[width=0.45\textwidth]{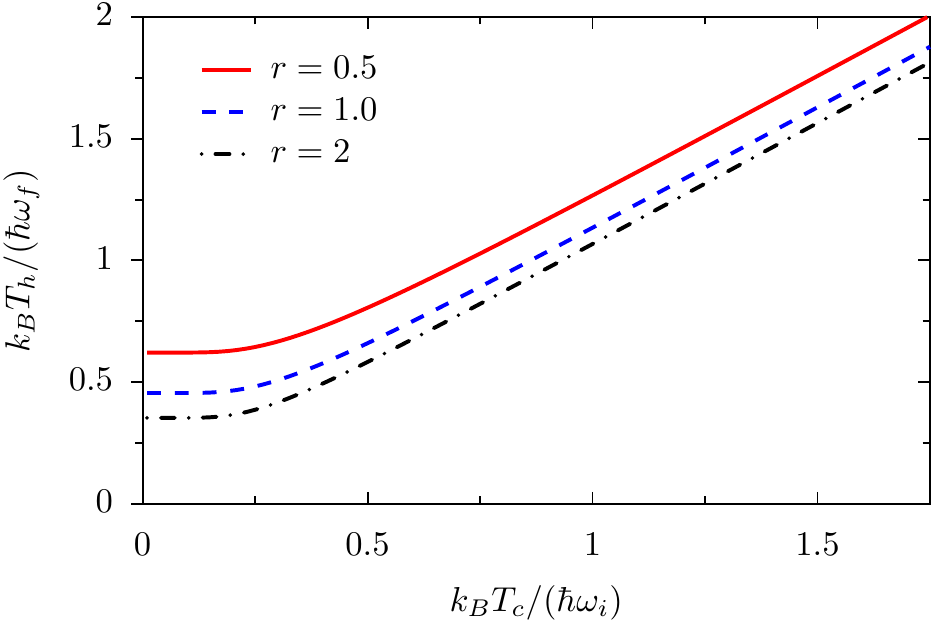}
\caption{ The hot temperature $T_h$ given by Eq. (\ref{bh}) of an Otto engine, which performs the same average work as a measurement driven engine with different measurement parameters $r$,  is displayed as a function of the temperature $T_c$ of the cold heat bath to which also the measurement engine couples in its final equilibration stroke. Both engines are operated with adiabatic work strokes and ideal thermalization. The scaling of the hot and cold temperatures with the frequencies in the compressed and expanded states, respectively, leads to a compression rate independent plot. The hot temperature displays a rapid transition between the asymptotic low-temperature  and the high-temperature behaviors $k_BT_h/(\hbar \o_f) \approx 1/\ln(1+8r)$ and $k_B T_h/(\hbar \o_f) \approx k_B T_c/(\hbar \o_i) +(4r-1)/(8r)$, respectively. }
\label{ThTc}
\end{figure}

For the comparison of a measurement driven engine and a conventional two temperature engine we consider work strokes with the same compression factor $\g$ , the same inverse initial temperature, i.e. $\b = \b_c$ and a bath at a high temperature chosen such that the average amounts of energies supplied to the working substances of both engines by the strokes between the states {\bf 1} and {\bf 2} agree with each other. The average energy imposed by the contact with a hot bath, given by eq. (\ref{Ebh}) must agree with 
  $\langle E_{\mathcal{M}} \rangle = \hbar \o_f /(8 r)$. This requirement  amounts for the inverse temperature of the hot heat bath to become 
\be
\b_h \hbar \o_f =\ln \left ( 1 + \frac{ 8r}{1 + 8 r \langle n \rangle^{(i)}_{\b_c}} \right )\:.
\ee{bh}
Figure~\ref{ThTc} exemplifies the resulting hot temperature $T_h$ as a function of the cold bath temperature $T_c$ for different measurement parameters $r$. The hot temperature is approximately independent of $T_  c$ at low temperatures and then turns over into an approximately linear dependence.   We note that due to the assumption of adiabatic work strokes with the supplied energy also the total work and the efficiency are equal for the Otto and the measurement energy.  
\begin{figure}
\includegraphics[width=0.45\textwidth]{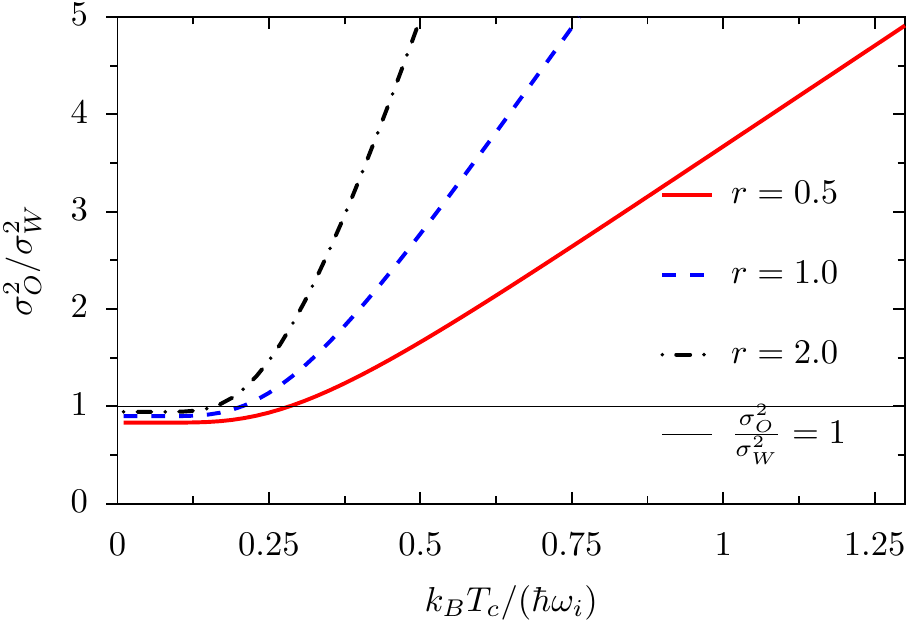}
\caption{The total work variances $\s^2_O$ of an Otto and $\s^2_W$ of a measurement engine are compared to each other as functions of the common temperature $T_c$ of the lower heat bath of the former and that of the measurement engine. The temperature of the hot heat bath of the Otto engine is chosen in dependence of the measurement parameter $r$ according to Eq. (\ref{bh}) so that the average total works of both engines agree with each other. The three displayed curves refer to different measurement parameters $r$. The work strokes of both engines are adiabatic and all thermalization strokes are perfect. In this case the considered ratio of work variances is independent of the compression ratio $\g$. The Otto engine performs slightly more reliably than the measurement driven engine only below the characteristic temperature $T^c_c$ given in eq. (\ref{Tcc}). Otherwise the measurement driven engine works better.  
 } 
\label{SOSM}
\end{figure}

Figure~\ref{SOSM} illustrates the ratio of the work variances characterizing an Otto and a measurement driven engine operated at equal average work output.
At the lowest temperatures the total work fluctuates less than a measurement driven engine which however becomes more reliable at slightly higher temperatures. The critical temperature $T^c_c$ above which the measurement engine performs more reliable than the Otto engine follows from the equations (\ref{s2W}) and (\ref{s2n}) in combination with eq. (\ref{bh}) as
\be
k_B T^c_c = \hbar \o_i/\ln(1+ 64 r^2 + 8 r \sqrt{2+r^2})\:.
\ee{Tcc}
The thermal energy at this critical temperature is always smaller than the level splitting of the work substance in the expanded state. It approaches this upper bound only in the limit of infinitely precise position measurements characterized by $r=0$ yielding a vanishing inverse hot temperature $\b_h =0$.
 
In the considered limit of adiabatic work strokes the variance ratio $\s^2_O/s^2_W$ is independent of the compression rate $\g$.

Finally, we note that, while a non-selective measurement always increases the energy of the harmonic oscillator working substance, in a conventional Otto engine, this is only the case  provided that the temperatures of the heat baths obey the inequality $\b_c > \g \b_h$. As a consequence, the efficiency $\h = 1 - \g^{-1}$ is always less than the Carnot efficiency $\h_{\text{Carnot}} = 1- \b_h/ \b_c$ \cite{KR}. The maximal efficiency is only attained   
for $\b_c = \g \b_h$ in which case, however, the average total work as well as the supplied energy vanish. In this particular situation, the 
working substance has already reached the Gibbs state at the higher temperature after completion of the adiabatic compression stroke and, on average, no energy uptake results from the contact with the hot heat bath.   
 We note that the fact that an adiabatic transformation of a Gibbs state exclusively passes through a series of Gibbs states parameterized by a running temperature is an exceptional property of working substances which allow uniform compression, such as the here considered harmonic oscillator. In the general case, when a parameter variation leads to a change of level distances of which some are not proportional to the other, an adiabatic transformation proceeds through non-equilibrium states and will typically also end in a non-equilibrium state.

\section{Conclusions}\label{conclusion}
We considered a microscopically small engine whose energy input results from a non-selective measurement of a working substance's observable that does not commute with the Hamiltonian of the working substance. During this measurement stroke, the working substance is supposed to be thermally isolated. The other strokes are the same as in a conventional Otto engine, consisting of compression before and expansion after the measurement stroke, and a final thermalization stroke by means of a weak contact with a heat bath  that allows the working substance to release energy and to come back to the state at which the cycle has started.  Also the work strokes (compression and expansion) are performed while the working substance is  thermally isolated. For the sake of concreteness we chose a harmonic oscillator with an externally controllable  frequency by which the compression and expansion strokes can be carried out. Further, we considered a non-selective Gaussian  measurement as the energy input stroke. 

In order to monitor the  energy profile of a cycle, the energy of the working substance is measured at the beginning of each stroke, resulting in four energy measurements per cycle. The resulting energy profile provides us with a rich diagnostic tool that not only yields the average behavior of the engine but also gives insight in its unavoidably random behavior. While the efficiency of the engine is universally given by $\h =1-\g^{-1}$ for adiabatic work processes it acquires an additional temperature and Husimi parameter  dependent factor multiplying the inverse compression parameter $\g^{-1}$. For non-adiabatic work strokes this factor is larger than one leading to the expected decrease of the efficiency for non-adiabatic work strokes. With increasing speed of the compression and expansion strokes the said factor grows until the average total work changes its sign and the device stops functioning as an engine.   

The energy supplied by a Gaussian position measurement is a random quantity with its most probable value at zero and a relatively small positive average value which only depends on the strength of the measurement but which is independent of the initial distribution of energy eigenstates.  Quite large fluctuations occur about the average value. The average but also the fluctuations increase with increasing measurement strength which can be quantified by the inverse measurement parameter $r$ defined in Eq. (\ref{r}) as the ratio of the variance of the measurement apparatus and the position variance of the oscillator in its ground state.

For adiabatic work strokes the total work is directly proportional to the measurement induced energy and hence inherits its fluctuations.  Additional randomness is added to the work when the work strokes are non-adiabatic. While the measurement parameter $r$ has a relatively small influence on the efficiency it strongly affects the reliability of the engine in that a more precise measurement (smaller measurement parameter $r$) improves the reliability. Likewise, the efficiency depends much weaker on the temperature than the reliability which decreases with increasing temperature. In contrast, the compression parameter influences the efficiency stronger than the reliability.

At the price of imperfect thermalization and non-adiabatic work strokes the engine can perform a cycle within a finite amount of time and hence can function with a non-vanishing power.
At short work stroke times $\t \o_i$ the power is negative, as is the efficiency, and then rapidly grows before it approaches zero in the adiabatic limit $\t \o_i \to \infty$. In between it displays several local maxima which are more pronounced at lower temperatures.

Finally we compared an Otto engine  with a measurement engine both of which are operated with adiabatic work strokes and perfect thermalization. In order to make the two devices comparable, the cold heat bath of the Otto engine has the same temperature as the bath of the measurement engine and the hot heat bath has a temperature that gives the same efficiency as the considered measurement engine has. This typically leads to a different reliability  which is better for the Otto engine only at rather low temperatures. At the still low temperatures where the thermal energy corresponds to the level distance of the working substance in the initial (expanded) state, the variance of the Otto engine rapidly becomes much larger than that of the measurement engine.

\appendix
\section{Measurement-induced transition probability}
We here detail two strategies to numerically evaluate the transition probabilities $T_{\mathcal{M}}(m,n)$, Eq. (\ref{TMmnM}), describing a Gaussian position measurement on a harmonic oscillator. 
\subsection{Gaussian representation}\label{TMG}
The Eq. (\ref{TMmnM}) can be expressed as an expectation value of a product of four  Hermite  polynomials with respect to the Gaussian probability density function with vanishing mean value and covariance matrix $\bM$ yielding
\be
\begin{split}
T_{\mathcal{M}}(m,n) &= \frac{2 \left (\det \bM \right )^{1/2}}{2^{n+m} n! m!} \langle H_n(\z) H_m(\z) H_n(\z') H_m(\z')  \rangle_{\bM}\\
&= \frac{2\p \left (\det \bM \right )^{1/2}}{2^{n+m} n! m!}\\
&\quad \times \sum_{i,j,k,l} H^i_n H^j_m H^k_n H^l_M \langle \z^{i+j} \z'^{k+l} \rangle_\bM\:,
\end{split}
\ee{TM2}  
where $\langle \bullet \rangle_\bM = \i \frac{d^2 \bfz}{2 \p \left ( \det \bM \right )^{1/2}} \bullet \exp \left \{-\frac{1}{2} \bfz \cdot \bM^{-1} \cdot \bfz \right \}$ is the said Gaussian average, and $H^k_m$ the $k$-th coefficient of the Hermite polynomial of the order $m$, $H_n(x) = \sum_{k=0}^n H^k_n x^k$. These coefficients can be obtained with the help of a symbolic computer system like Mathematica or Maple.   
The expectation values of the monomials $\z^k \z'^l$ follow from the Gaussian recursion relations 
\be
\begin{split}
\langle \z^k \z'^l \rangle_\bM  &= (k-1) \langle \z^2 \rangle_\bM \langle \z^{k-2} \z'^l\rangle_\bM\\
&\quad  + l \langle \z \z' \rangle_\bM \langle \z^{k-1} \z'^{l-1}\rangle_\bM\\
&=k \langle \z \z' \rangle_\bM \langle \z^{k-1} \z'^{l-1}\rangle_\bM\\
&\quad + (l-1) \langle \z'^2 \rangle_\bM \langle \z^k \z'^{l-2} \rangle_\bM \:.
\end{split}
\ee{gr}
The second moments are given by the covariance matrix $\bM$ to read
\be
\begin{split}
\langle \z^2 \rangle_\bM = \langle \z'^2 \rangle_\bM& = \frac{8 r +1}{4(4 r +1)}\\
\langle \z \z' \rangle_\bM &=\frac{1}{4(4 r +1)}\:.
\end{split}
\ee{zr} 
This approach is limited by the fact that the terms in the summands  on the right hand side of Eq. (\ref{TM2}),  with growing $n$ and $m$, assume  exceedingly large positive and negative values which eventually prohibit a reliable calculation of the sum. 
\subsection{Generating function}\label{TMgf}
In order to make use of the generating function of the Hermite polynomials, \cite{AS}, reading,  
\be
\sum_{n=0}^\infty \frac{u^n}{n!} H_n(x) = e^{2 x u -u^2}
\ee{hgf}
we introduce the auxiliary quantities $R^{m' n'}_{m n}$ defined as
\be 
\begin{split}
R^{m' n'}_{m n} &= \i d\z d\z' H_n(\z) H_m(\z) H_{n'}(\z') H_{m'}(\z') \\
& \quad \times  e^{-\z^2 +\z'^2} e^{-(\z -\z')^2/(8 r)}  \; .
\end{split}
\ee{R}
The transition probability (\ref{TMmnM}) caused by the measurement can then be expressed as
\be
T_\bM(m,n) = \frac{1}{2^{m+n} n! m! \p} R^{m,n}_{m,n}\:.
\ee{TR}
Introducing the generating function 
\be
\X(u,v,w,z) = \sum_{m,n,m',n'} \frac{u^n v^{n'} w^m z^{m'}}{n! m! n'! m'!} R^{m',n'}_{m,n}\:,
\ee{fR}
where the sum is taken from $0$ to $\infty$ with respect to all indices, one can write the auxiliary coefficients in terms of according  derivatives
as
\be
R^{m' n'}_{m n} = \left .\frac{\partial^{n+n'+m+m'} }{\partial u^n \partial v^{n'} \partial w^m \partial z^{m'}} \X(u,v,w,z)  \right \vert_{\substack{u=v=w\\=z=0}}  \; .
\ee{Rf}
Using the definition (\ref{fR}) one can express the generating function $\X(u,v,w,z)$ with the help of Eq. (\ref{hgf}) in terms of a Gaussian double integral which can be performed yielding
\be
\X(u,v,w,z) = 2 \p \sqrt{\frac{r}{1+4 r}} e^{\c(u,v,w,z)}\:,
\ee{Xc}
where 
\begin{widetext}
\be 
\c(u,v,w,z) = \frac{-1}{1+4 r} \left [ \frac{1}{2} (u^2\! +\! v^2\! +w^2\! +\!z^2) -(u\!+\!w)(z\!+\!v) - (1+8r) (uw \!+\! vz) \right ]\:.
\ee{c}  
\end{widetext}
The derivatives of $\X(u,v,w,z)$ can be determined with the help of a symbolic computer language up to a maximal order depending on the RAM size of the used computer and on the magnitude of the parameter $r$.    
\section{Average energy supplied by a position measurement}\label{avdn}
The average energy change of a particle moving in a one-dimensional potential $V(q)$ caused by a generalized Gaussian position measurement 
in a state characterized by the density matrix $\r$ can be expressed as
\be
\begin{split}
\langle E_{\mathcal{M}} \rangle &= \Tr H \F^{\mathcal{M}}(\bar{\r}) - \Tr H \bar{\r}\\
&= \Tr \left (\F^{\mathcal{M}}(H) - H \right ) \bar{\r}\:,
\end{split}
\ee{avEM}
where $H= p^2/(2m) + V(q)$ is the Hamiltonian of the particle and the measurement operation $\F^{\mathcal{M}}$ is given by Eq. (\ref{FM}). Here $\bar{\r}$ is the diagonal part of the density matrix $\r$ with respect to the energy basis as it results from the energy measurement prior to the position measurement, \cite{THM}. In going to the second line of the above equation we used the fact that $\F^{\mathcal{M}}$ is unital. Moreover this measurement operation leaves the potential part of the Hamiltonian unchanged which therefore cancels resulting in:
\be
\begin{split}
\langle E_{\mathcal{M}} \rangle &= \frac{1}{2m} \Tr (\F^{\mathcal{M}}(p^2) - p^2)\bar{\r}\\
&= \frac{1}{2m} \Tr \left (\i \frac{dx}{\sqrt{2 \p \s^2}} e^{-(q-x)^2/(2 \s^2)} p^2\left (\frac{1}{4 \s^2} \right ) - p^2 \right ) \bar{\r}\:.
\end{split}
\ee{EMp}    
In going to the second line we introduced the transformed momentum operator 
$p(t)$ given by
\be
p(t) = e^{(q-x)^2 t} p e^{-(q-x)^2 t}
\ee{pte}
that can readily be expressed as 
\be
p(t) = p + 2 i \hbar t (q-x)\:.
\ee{pt}
Hence one finds
\be
p^2\left (\frac{1}{4 \s^2} \right ) = p^2 +i \frac{\hbar}{\s^2} (q-x) p + \frac{\hbar}{2 \s^2} - \frac{\hbar^2}{4 \s^4} (q-x)^2\:.
\ee{p2} 
Now the Gaussian integration over $x$ can be performed. Noting that, with the Gaussian $x$-integral being centered at $x=q$, the  term linear in $(q-x)$ vanishes and the quadratic contribution $(q-x)^2$ yields $\s^2$ one finally obtains 
\be
\langle E_{\mathcal{M}} \rangle = \frac{\hbar^2}{8 m \s^2}
\ee{EMs}
in accordance with the result given in Ref.~\cite{YTK}. Most notably, the average measurement supplied energy is positive and independent of the potential $V(q)$ and of the  density matrix $\r$ characterizing the state of the considered particle immediately before the measurement sequence of energy, position and again energy. 
\section{Averages of work and supplied energy} \label{Awe}
According to Eq. (\ref{W}) the average work consists of four additive contributions given by
\be
\langle W \rangle = \hbar \o_i (\langle l \rangle_3 - \langle n \rangle_0) + \hbar \o_f (\langle m \rangle_1 -\langle k \rangle_2)
\ee{AW}
and the average supplied energy accordingly becomes
\be
\langle E_{\mathcal{M}} \rangle = \hbar \o_f \left ( \langle k \rangle_2 - \langle m \rangle_1 \right )\:,
\ee{AEM} 
where 
\begin{align}
\langle n \rangle_0 &= \sum_n n p(n,\th) \label{n0} \:,\\
\langle m \rangle_1 &=\sum_{m,n} m T(m,n) p(n,\th) \label{m1}\:,\\
\langle k \rangle_2 &=\sum_{k,m,n} k T_\mathcal{M}(k,m) T(m,n) p(n,\th) \label{k2}\:,\\
\langle  l \rangle_3 &= \sum_{l,k,m,n} l T(l,k) T_{\mathcal{M}}(k,m) T(m,n) p(n,\th)\:. \label{l3}
\end{align}   
Combining eqs. (\ref{EMs}) and (\ref{AEM}) one finds for the  populations difference between {\bf 2} and {\bf 1} the following expression
\be
\langle z \rangle =\langle k \rangle_2 - \langle m \rangle_1 = \frac{1}{8 r}\:,
\ee{kmr}
which only depends on the measurement-uncertainty parameter $r$.


\begin{thebibliography}{99}
\bibitem{note} Note, that in Ref.~\cite{YTK} the inverse is introduced as the compression rate.
\end{thebibliography}
\end{document}